\documentclass[english,twocolumn,pra,superscriptaddress,showpacs]{revtex4}

\usepackage{graphicx}
\usepackage{epsfig,psfrag}

\usepackage[dvips]{color}
\definecolor{darkgreen}{rgb}{0,.6,0}

\newcommand{\red}{\color{red}}

\begin{document}

\title{Electron penetration in the nucleus and its effect on the quadrupole
  interaction}

\author{Katrin Koch}
\email{Katrin.Koch@cpfs.mpg.de}
\affiliation{Max Planck Institute for Chemical Physics of Solids,
  N\"othnitzer Str. 40, \\ DE-01187 Dresden, Germany}

\author{Klaus Koepernik}
\affiliation{IFW Dresden, Institute for Solid State Research,
  P.O.~Box 270116, \\ DE-01171 Dresden, Germany}

\author{Dimitri Van Neck}
\affiliation{Center for Molecular Modeling, Ghent University,
  Proeftuinstraat 86, \\ BE-9000 Ghent, Belgium}

\author{Helge Rosner}
\affiliation{Max Planck Institute for Chemical Physics of Solids,
  N\"othnitzer Str. 40, \\ DE-01187 Dresden, Germany}

\author{Stefaan Cottenier}
\email{Stefaan.Cottenier@ugent.be}
\affiliation{Center for Molecular Modeling, Ghent University,
  Proeftuinstraat 86, \\ BE-9000 Ghent, Belgium}
\affiliation{Instituut voor Kern- en Stralingsfysica and INPAC,
  K.U.Leuven, Celestijnenlaan 200 D, BE-3001 Leuven, Belgium}

%
%
%
%
%
%
%
%
%
%
%

\begin{abstract}
A series expansion of the interaction between a nucleus and its
surrounding electron distribution provides terms that are well-known
in the study of hyperfine interactions: the familiar quadrupole
interaction and the less familiar hexadecapole interaction. If the
penetration of electrons into the nucleus is taken into account,
various corrections to these multipole interactions appear. The best
known one is a scalar correction related to the isotope shift and the
isomer shift. This paper discusses a related tensor correction, which
modifies the quadrupole interaction if electrons penetrate the
nucleus: the quadrupole shift. We describe the mathematical formalism
and provide {first-principles} calculations of the quadrupole shift
for a large set of solids. Fully relativistic calculations that
explicitly take a finite nucleus into account turn out to be
mandatory. Our analysis shows that the quadrupole shift becomes
appreciably large for heavy elements.  Implications for experimental
high-precision studies of quadrupole interactions and quadrupole
moment ratios are discussed.
A literature review of other small quadrupole-like effects is
presented as well.
\end{abstract}

\date{\today}


\maketitle

%
%
%
%
%
%
%
%
%
%
%

\section{Introduction}

Atomic nuclei are no mathematical point charges, but objects with a
shape and a size. This affects the way in which they interact with
electrons, especially when electrons penetrate the nuclear volume and
render the usual `far-field' approximation invalid. These `near-field'
effects lead to tiny corrections to all terms in the multipole
expansion for the electrostatic interaction between nuclei and
electrons. The correction to the monopole term corresponds to
experimentally well-known phenomena: the isotope shift in atomic
spectroscopy and the isomer shift in M\"ossbauer spectroscopy. An
analogous correction to the quadrupole term -- coined here the {\it
  quadrupole shift}~\footnote{The term ``quadrupole shift" can appear
  in the literature with another meaning than the one defined in this
  paper. In NQR, it sometimes refers to the anisotropy of the
  quadrupole interaction with the applied field~\cite{QSotherName1},
  while in high-precision optical spectroscopy it refers to the
  interaction between an atom's nucleus and the background
  electric-field gradient provided by an ion trap~\cite{QSotherName2}
  .} (QS) -- should exist as well. The existence of such an effect has
been touched upon a few times in the literature of the past
decades~\cite{Pyykko1970b,steffen1, steffen2, Band1985, Thyssen,
  VanStralen2003}, but to our knowledge a systematic study is
lacking. In this paper, we present a mathematical treatment of the
quadrupole shift by a twofold application of first-order perturbation
theory, which leads to a simple analytical expression
(Secs.~\ref{sec-formalism}-\ref{subsec-overlap-consequences}).  We
point out that in order to compute numerical values for the quadrupole
shift from {first-principles}, it is necessary to perform fully
relativistic calculations that take explicitly a finite nucleus into
account (Sec.~\ref{sec-comp-aspects}). Density Functional Theory
calculations of the quadrupole shift for a set of simple crystals show
that the size of the quadrupole shift strongly grows with the mass of
the isotope, an effect that turns out to have an electronic rather
than a nuclear origin (Sec.~\ref{sec-implementation-trends}). Except
for the heaviest elements (actinides), the quadrupole shift is only a
minor correction to the quadrupole interaction. We discuss how it
shows up in experiments, and how it could possibly be exploited to
improve the accuracy of experimentally determined quadrupole moments
(Sec.~\ref{sec-exp-impl}).
Especially for the experimental determination of ratios between
nuclear electric quadrupole moments, the accuracy that can be reached
by the most precise molecular beam spectroscopy experiments is good
enough to make it relevant in some cases to take quadrupole shift
corrections into account. The quadrupole shift is only one of a set of
small effects that can affect the regular quadrupole interaction. The
(sometimes fairly old) literature on these other effects is reviewed
in App.~\ref{appendix}. We suggest that for high-precision studies it is
relevant to revisit these small quadrupole-like perturbations with
modern computational methods.

%
%
%
%
%
%
%
%
%
%
%

\section{Formalism}
\label{sec-formalism}

\subsection{Classical interaction energy without charge-charge overlap}
\label{subsec-no-overlap}
The {\it classical} electrostatic interaction energy between a
positive (nuclear) charge distribution $\rho(\vec r)$ and a potential
$v(\vec r)$ due to a surrounding (electron) charge distribution
$n(\vec r^{\,\prime})$ is formally given by (with $\epsilon_0$ being
the electric constant)
\begin{eqnarray}
\label{Eint1}
E = \int \rho(\vec r)v(\vec r)d\vec r
=  \frac{1}{4\pi\epsilon_0}\int \int \frac{\rho(\vec r)
n(\vec r^{\,\prime})}{|\vec r
- \vec r^{\,\prime}|}d\vec r d\vec r^{\,\prime},
\end{eqnarray}
and can be expressed by the standard multipole expansion in spherical
harmonics \cite{abragam}:
\begin{eqnarray}
\label{laplexp}
\frac{1}{|\vec r - \vec r^{\,\prime}|}
=\sum_{l,m}\frac{4\pi}{2l+1}\frac{r_<^l}{r_>^{l+1}}
Y_{l,m}^*(\Omega) Y_{l,m}(\Omega{^{\prime}}),
\end{eqnarray}
with $r_<=\min(r,r^{\,\prime})$ and $r_>=\max(r,r^{\,\prime})$. This
leads to an infinite sum of double integrals, each with the dimension
of energy:
\begin{eqnarray}
\label{eq-sumofEi}
E & = & \sum_{l\!=\!0}^{\infty} E_{2l} \ = \ E_0 + E_2 + E_4 + \ldots
\end{eqnarray}
(Odd terms will vanish in the cases of interest here, see
Sec.~\ref{subsec-with-overlap}.) It is the second term $E_2$ that will
be of interest in the present work:
\begin{widetext}
\begin{eqnarray}
E_2 & = & h \nu_Q
\ = \  \frac{1}{4\pi\epsilon_0} \frac{4\pi}{5} \sum_{m=-2}^{+2}
\int \int
\rho(\vec r) n(\vec r^{\,\prime}) \frac{r_<^2}{r_>^{3}}
Y_{2,m}^*(\Omega) Y_{2,m}(\Omega{^{\,\prime}})d\vec r d\vec
r^{\,\prime}.
\label{eq-E2-correct}
\end{eqnarray}
\end{widetext}
The frequency $\nu_Q$ is experimentally accessible, and is called the
{\it nuclear quadrupole coupling constant} (NQCC). Due to the varying
assignment of $r_<$ and $r_>$ to `nuclear' ($r$) or `electron'
($r^{\,\prime}$) coordinates, quantities as $E_2$ are an intricate
mixture of properties of both charge distributions $\rho(\vec r)$
and $n(\vec r)$. Only in the special case where both charge
distributions do not overlap ($r_<\equiv r$ and $r_>\equiv
r^{\,\prime}$), Eq.~(\ref{Eint1}) can be written in terms of properties
that depend entirely on only one of the charge distributions:
\begin{eqnarray}
\label{Eint2}
E=\sum_{l,m}Q_{lm}^*V_{lm} \ ,
\end{eqnarray}
where $ Q_{lm}$ and $V_{lm}$ are the components of the nuclear
multipole moment and electric multipole field tensors of rank $l$,
respectively:
\begin{eqnarray}
\label{Qlm}
 Q_{lm}&=&\sqrt{\frac{4\pi}{2l+1}}\int r^l\rho(\vec r)
Y_{lm}(\Omega)d\vec r
\\
 V_{lm}&=& \frac{1}{4\pi\epsilon_0}\sqrt{\frac{4\pi}{2l+1}}\int \frac{1}{r^{\,\prime l+1}}
n(\vec r^{\,\prime})Y_{lm}(\Omega^{\,\prime})d\vec r^{\,\prime}.
\label{Vlm}
\end{eqnarray}
When this formalism is applied to describe nuclei and electrons, the
simplification by Eq.~(\ref{Eint2}) can never be made: $s$-electrons
and relativistic $p_{\frac{1}{2}}$-electrons have a non-zero
probability to appear at $r\!=\!0$, and therefore the nuclear and
electron charge distributions always overlap.
Nevertheless, motivated by the very small size of the region where
this overlap happens compared to the volume of the rest of the atom,
one can in a first approximation neglect this concern and apply
Eq.~(\ref{Eint2}) to atoms, molecules and solids. This is where the
concept originates of an electric-field gradient (EFG) tensor
($V_{2m}$) that interacts with a nuclear quadrupole moment tensor
($Q_{2m}$) to produce an experimentally observable interaction energy
($E_2$). Although $E_2$ itself is a well-defined observable property,
its description by a quadrupole interaction energy only
\begin{eqnarray}
\label{eq-E2-approx}
E_2 & \approx & \sum_{m=-2}^{+2} Q_{2m}^*V_{2m}
\end{eqnarray}
rather than by Eq.~(\ref{eq-E2-correct}) is an approximation.

%
%
%
%
%
%
%
%
%
%
%

\subsection{Overlap corrections}
\label{subsec-with-overlap}

We will now derive explicit expressions for the corrections that need
to be added to Eq.~(\ref{eq-E2-approx}) to obtain
Eq.~(\ref{eq-E2-correct}) (and similarly for other values of
$l$). Rather than using the multipole expansion in spherical harmonics
from Eq.~(\ref{laplexp}), we start from a Taylor expansion of the
electrostatic potential $v(\vec r)={1}/({4\pi\epsilon_0})\int n(\vec
  r^{\,\prime})/|\vec r-\vec r^{\,\prime}| \ d \vec r^{\,\prime}$ in
  the interaction energy of Eq.~(\ref{Eint1}):
%
\begin{eqnarray}
\nonumber
E &=& \int \rho (\vec r) v(\vec r) d \vec r
=v(0)\int \rho (\vec r) d \vec r
\nonumber
\\
&&
+ \sum_{i}v_i(0) \int x_i \rho(\vec r) d \vec r \nonumber
\\
&&
+\frac{1}{2!} \sum_{i,j}v_{ij}(0) \int x_i x_j \rho (\vec r) d \vec r
\nonumber
\\[.5cm]
&&
+\frac{1}{3!} \sum_{i,j,k}v_{ijk}(0) \int x_i x_j x_k\rho (\vec r)
d \vec r \nonumber
\\
&&
+\frac{1}{4!} \sum_{i,j,k,l}v_{ijkl}(0) \int x_ix_jx_kx_l\rho (\vec r)
d \vec r
+\mathcal{O}(6). \label{eq-Etot-Taylor}
\end{eqnarray}
%
In order to recognize in this expression the multipole moments and
multipole fields from Eq.~(\ref{Qlm}) and Eq.~(\ref{Vlm}), one has to
make substitutions like this one (the example is for the quadrupole moment):
%
\begin{eqnarray}
\nonumber
\int x_i x_j \rho (\vec r) d \vec r & = & \frac{1}{3}
\underbrace{\int (3x_i x_j-r^2\delta_{ij})\rho (\vec r) d \vec r}
_{Q_{ij}}
\\
&&
\ + \ \frac{1}{3}\int r^2 \rho(\vec r)d\vec r\ \delta_{ij} \ ,
 \label{eq-Q2-substitution}
\end{eqnarray}
%
where $Q_{ij}$ are the components of the quadrupole tensor $Q_{2m}$
(Eq.~(\ref{Qlm})), but now in Cartesian form.  This yields for the
first three even orders in Eq.~(\ref{eq-Etot-Taylor}) the following
nuclear multipole moments in Cartesian form:
\begin{eqnarray}
\label{M}
M&=&\int \rho (\vec r) v(\vec r) d \vec r=eZ
\\
Q_{ij}&=&\int (3x_ix_i-r^2\delta_{ij}) \rho(\vec r) d \vec r
\\
\nonumber
H_{ijkl}&=&\int 3\cdot5 (7x_ix_jx_kx_l
\\
\label{H}
&&
 \ \ - f^H(x_i,x_j,x_k,x_l)) \rho(\vec r) d \vec r
\end{eqnarray}
with $f^H(x_i,x_j,x_k,x_l)=r^2\Big[
 x_ix_j\delta_{kl}
+x_ix_k\delta_{jl}
+x_ix_l\delta_{kj}
+x_jx_k\delta_{il}
+x_jx_l\delta_{ik}
+x_kx_l\delta_{ij}
\Big]
- \frac{r^4}{5} \Big[
 \delta_{ij}\delta_{kl}
+\delta_{ik}\delta_{jl}
+\delta_{il}\delta_{jk}
\Big]$.
\\
The corresponding electric multipole fields in Cartesian form are:
\begin{eqnarray}
\label{V}
V&=&v(0)
\\
V_{ij}&=&(\partial_i\partial_j v(0) - \frac{1}{3}\Delta \delta_{ij} )
\Delta v(0)
\\
V_{ijkl}&=&\partial_i\partial_j\partial_k\partial_lv(0)-f^V_{ijkl}
\Delta v(0)
\label{Vijkl}
\end{eqnarray}
with
$f^V_{ijkl}=\Big[ \partial_i\partial_j\delta_{kl}
  +\partial_i\partial_k\delta_{jl} +\partial_i\partial_l\delta_{kj}
  +\partial_j\partial_k\delta_{il} +\partial_j\partial_l\delta_{ik}
  +\partial_k\partial_l\delta_{ij} \Big] - \frac{\Delta}{5} \Big[
  \delta_{ij}\delta_{kl} +\delta_{ik}\delta_{jl}
  +\delta_{il}\delta_{jk} \Big]$.
The expressions in Eqs.~(\ref{M}) to (\ref{H}) and Eqs.~(\ref{V}) to
(\ref{Vijkl}) are identical to the ones in Eqs.~(\ref{Qlm})
and~(\ref{Vlm}), respectively. They have the same number of degrees of
freedom: 1, 5 and 9 for the zeroth, second and fourth order
moment/field.

After having inserted into Eq.~(\ref{eq-Etot-Taylor}) all
substitutions as in Eq.~(\ref{eq-Q2-substitution}), the interaction
energy can be written as
%
\begin{eqnarray}
\nonumber
E&=&
\underbrace{M\cdot V}_{\mbox{MI}}
\ + \ \underbrace{\frac{1}{3!}\{ r^2 \} \Delta v(0)}_{\mbox{MS$^{(1)}$}}
\ + \ \underbrace{\frac{1}{5!}\{r^4\} \Delta^2 v(0)}_{\mbox{MS$^{(2)}$}}
\\
\nonumber
&&
\ + \ \underbrace{\frac{1}{2!}\frac{1}{3}\sum_{ij}Q_{ij}V_{ij}}_{\mbox{QI}}
\\
\nonumber
&&
\ + \ \underbrace{\frac{1}{28}\sum_{ij}\{(x_ix_j-\frac{r^2}{3}\delta_{ij})r^2\}
(\partial_i\partial_j-\frac{\Delta}{3}\delta_{ij})\Delta v(0)}_
{\mbox{QS$^{(1)}$}}
\\
&&
\ + \ \underbrace{\frac{1}{4!}\frac{1}{105}\sum_{ijkl}H_{ijkl}V_{ijkl}}_
{\mbox{HDI}}
\ + \ \mathcal{O}(6),
\label{generalE}
\end{eqnarray}
%
where all integrations over the nuclear charge density $\rho(\vec r)$
are noted in short-hand by \{curled brackets\}. Eq.~(\ref{generalE})
contains no odd order terms (dipole, octupole,$\ldots$), since nuclei
have no odd order electric moments due to time reversal
symmetry~\cite{steffen2}.  We see that Eq.~(\ref{generalE})
contains dot products between multipole moments and fields as in
Eq.~(\ref{Eint2}): the monopole (MI), quadrupole (QI), hexadecapole
(HDI),$\ldots$ interactions. These are the only contributions in the
case without charge-charge overlap. Additionally, an infinite set of
even order correction terms appears now as well -- due to parity,
there are no odd order corrections. In Tab.~\ref{corTab}, a general
naming system and a corresponding set of symbols are presented: the
$n^{th}$ order {\it quasi} multipole {\it moment} multiplied (dot
product) with the $n^{th}$ order {\it quasi} multipole {\it field}
leads to the $n^{th}$ order multipole {\it shift}. From the general
trends in this table one can infer the structure of the higher order
corrections that were not explicitly derived in Eq.~(\ref{generalE})
-- they are shown in the table in red.

\begin{table*}[t]
\caption {Systematic overview of nuclear multipole and quasi multipole
  moments and electric multipole and quasi multipole fields that
  appear in the multipole expansion of two interacting (and
  overlapping) classical charge distributions.
  The first column gives the regular multipole expansion for point
  nuclei: the monopole, quadrupole and hexadecapole interactions.
  The next columns give the quasi multipole moments/fields for every
  multipole interaction, denoted by a tilde: these are corrections to
  the multipole interactions due to electron penetration into an
  extended nucleus.
  Colored text is by generalization only, and is not systematically
  derived in this work. The objects in each line are spherical tensors
  of a given rank (rank 0 for line 1, rank 2 for line 2, rank 4 for
  line 3, $\ldots$).}
\label{corTab}
\begin{center}
\begin{tabular*}{\linewidth}{c@{\extracolsep\fill}cccc}
\hline
\hline
\\[-0.25cm]
Order & Multipole &  First order  &  Second order  &  $\dots$
\\
 & moment  & quasi moment & quasi moment &
\\
& / field & / quasi field & / quasi field &
\\[.15cm]
\hline
\\[-0.35cm]
$\mathcal{O}(0)$
&
\begin{tabular}{c}
MI:
\\
$M\propto r^0Y_{00}$
\\
$V\propto v(0)$
\end{tabular}
&
\begin{tabular}{c}
MS$^{(1)}$:
\\
$\tilde M^{(1)}\propto \{r^2 Y_{00}\}$
\\
$\tilde V^{(1)}\propto \Delta v(0)$
\end{tabular}
&
\begin{tabular}{c}
 MS$^{(2)}$:
\\
$\tilde M^{(2)} \propto \{r^4 Y_{00}\}$
\\
$\tilde V^{(2)}\propto \Delta^2 v(0)$
\end{tabular}
& $\dots$
\\
\\[.1cm]
$\mathcal{O}(2)$
&
\begin{tabular}{c}
QI:
\\
$Q\propto r^2Y_{20}$
\\
$V_{ij}\propto \partial_{ij} v(0)$
\end{tabular}
&
\begin{tabular}{c}
QS$^{(1)}$:
\\
$\tilde Q^{(1)}\propto\{r^4 Y_{20}\}$
\\
$\tilde V_{ij}^{(1)}\propto\partial_{ij}\Delta v(0)$
\end{tabular}
&
\begin{tabular}{c}
\red QS$^{(2)}$:
\\
\red $\tilde Q^{(2)}\propto\{r^6 Y_{20}\}$
\\
\red $\tilde V_{ij}^{(2)}\propto\partial_{ij}\Delta^2 v(0)$
\end{tabular}
& $\dots$
\\
\\[.1cm]
$\mathcal{O}(4)$
&
\begin{tabular}{c}
HDI:
\\
$H\propto r^4Y_{40}$
\\
$V_{ijkl}\propto \partial_{ijkl}v(0)$
\end{tabular}
&
\begin{tabular}{c}
\red HDS$^{(1)}$:
\\
\red $\tilde H^{(1)}\propto\{r^6 Y_{40}\}$
\\
\red $\tilde V_{ijkl}^{(1)}\propto \partial_{ijkl}\Delta v(0)$
\end{tabular}
&
\begin{tabular}{c}
\red HDS$^{(2)}$:
\\
\red $\tilde H^{(2)}\propto\{r^8 Y_{40}\}$
\\
\red $\tilde V_{ijkl}^{(2)}\propto\partial_{ijkl}\Delta^2 v(0)$
\end{tabular}
& $\dots$
\\
\\[.1cm]
$\dots$& $\dots$  &  $\dots$ &  $\dots$& $\dots$
\\[.05cm]
\hline
\hline
\end{tabular*}
\end{center}
\end{table*}

There is a qualitative difference between the multipole fields in the
first column of Tab.~\ref{corTab} and the quasi multipole fields in
all other columns. The multipole fields depend on the potential $v(0)$
at the nucleus, which depends via integration on the charge
distribution everywhere else in the system. Multipole fields are
therefore integrated quantities, determined by the entire density. The
quasi multipole fields depend on the Laplacian of the potential at the
nucleus ($\Delta v(0)$), which is by the Poisson equation proportional
to the electron charge density at the nucleus ($n(0)$) ($\Delta v(0)
=- n(0)/\epsilon_0$). Quasi multipole fields are therefore point
quantities, determined by the electron density in a single point only.


In the next section, the results of Eq.~(\ref{generalE}) and
Tab.~\ref{corTab} for a system of two classical charge distributions
will be translated to a quantum formulation. This will make it
applicable to atoms, molecules and solids. Known experimental
consequences of these overlap correction terms will be summarized in
Sec.~\ref{subsec-overlap-consequences}. The core of the present work
deals with the first order quadrupole shift QS$^{(1)}$, which is the
first order correction to the quadrupole interaction.

%
%
%
%
%
%
%
%
%
%
%

\narrowtext

\subsection{Quantum formulation}\label{Hamiltonians}

In order to translate Eq.~(\ref{generalE}) to quantum mechanics,
Hamiltonian operators corresponding to all its terms are required. The
structure of Eq.~(\ref{generalE}) suggests a perturbation theory
treatment, with the monopole interaction term as the unperturbed
Hamiltonian, and the other terms as small perturbations. The monopole
term depends via $r^0$ on the (small) nuclear coordinate ($r \propto
10^{-15}$~m) and via $1/r^{\,\prime}$ on the electronic coordinate
($r^{\,\prime} \propto 10^{-10}$~m). Among all small corrections in
Tab.~\ref{corTab}, the two largest ones are the quadrupole interaction QI
and the first order monopole shift MS$^{(1)}$ -- both have a $r^2$ in
their nuclear parts and a second derivative of the electrostatic
potential (leading to $1/r^{\,\prime 3}$) in their electronic
parts. These two leading corrections will be taken as the small
perturbation.

The Hamiltonians that correspond to the entries in Tab.~\ref{corTab}
operate on the direct product space of wave functions for the nuclear
and the electron subspaces. The ground state of the monopole
Hamiltonian is a direct product between the nuclear ground state and
the electronic ground state wave function. With
$\hat{M}=eZ\hat{1\!\mbox{l}}$ (Eq.~(\ref{M}), $\hat{1\!\mbox{l}}$ is
the identity operator on the nuclear space) and $\hat{V}=\hat{v}(0)$
(Eq.~(\ref{V}), $\hat{v}(0)$ is an operator on the electronic space
that returns the potential at $\vec{r}\!=\!\vec{0}$ due to a given
wave function $\Psi$), the unperturbed monopole interaction
Hamiltonian is
\begin{eqnarray}
\label{MIH}
\hat{\mathcal{H}}_{MI} =   eZ \ \hat{1\!\mbox{l}} \otimes \hat{v}(0) .
\end{eqnarray}
Evaluating this for the ground state wave function $\left|I \otimes
\Psi_0\right>$ of the combined nuclear+electronic system leads to
($\left|I\right>$ is the ground state of the nucleus, and
$\left|\Psi_0\right>$ the ground state of the electron system with a
point nucleus):
\begin{eqnarray}
E_0^{pn} & = & \left<\Psi_0 \otimes I \right| \hat{\mathcal{H}}_{MI}
\left| I \otimes |\Psi_0\right>
\nonumber
\\ & = & \left<I\right| eZ
\ \hat{1\!\mbox{l}} \left|I\right> \cdot \left<\Psi_0 \right|
\hat{v}(0) \left|\Psi_0\right>
\nonumber
\\ & = & eZv(0) \ ,
\label{eq-E0}
\end{eqnarray}
which is the leading term in Eqs.~(\ref{Eint2})
or~(\ref{generalE}). The label {\it pn} (`point nucleus') emphasizes
the difference with $E_0$ from Eq.~(\ref{eq-sumofEi}). The quantity
$v(0)$ -- the electrostatic potential at the nuclear site for a point
nucleus -- is accessible by {first-principles} codes.

The perturbing Hamiltonian is (see Tab.~\ref{corTab} for the notation):
\begin{eqnarray}
\label{eq-Hperturb}
\hat{\mathcal{H}}_P & = & \hat{\mathcal{H}}_{QI} + \hat{\mathcal{H}}_{{MS}^{(1)}}.
\end{eqnarray}
In {first order perturbation theory}, the energy corrections due
to this perturbation are found by evaluating the perturbing
Hamiltonian in the ground state of the unperturbed
Hamiltonian. Assuming a non-degenerate ground state in the electron
subspace, it is advantageous to write the Hamiltonians immediately in
a more familiar form where the electronic matrix elements are already
evaluated and are treated as known (=computable) quantities. After
similar algebra as for the monopole Hamiltonian, this leads to this
form for the monopole shift Hamiltonian (it contains the mean square
radius $\langle r^2\rangle$ of the nucleus and the electron density
$n(0)$ at the position of the nucleus):
\begin{eqnarray}
\label{MSH}
\hat \mathcal{H}_{{MS}^{(1)}} = -\frac{eZ}{6\epsilon_0}
n(0) \ \langle r^2\rangle  \hat{1\!\mbox{l}}.
\end{eqnarray}
The quadrupole Hamiltonian $\hat \mathcal{H}_{QI}$ contains the
(spectroscopic) quadrupole moment of the nucleus $Q$ and the
quadrupole field of the electrons $V_{zz}$ (principle component of the
electric-field gradient tensor) (see e.g.~Ref.\cite{abragam}):
\begin{eqnarray}
\hat \mathcal{H}_{QI} &=&
\frac{{ e Q} { V_{zz}}}{4(2I-1)I\hbar^2}
\left[\left(3\hat I_z^2 -\hat I^2\right)
+ \frac{1}{2}\eta\left(\hat I_+^2 + \hat I_-^2 \right)
\right].
\nonumber
\\
\label{eq-regular-QI-hamiltonian}
\end{eqnarray}
Diagonalizing these two Hamiltonians in the nuclear states leads to
the desired energy corrections in first order perturbation. Formally:
\begin{eqnarray}
E^{[1]} & = & E_0^{pn} \ + \ \langle I |\hat
\mathcal{H}_{{MS}^{(1)}}+\hat \mathcal{H}_{QI}
|I \rangle
\nonumber
\\
&  = & E_0^{pn} \ + \
\langle I| \hat \mathcal{H}_{{MS}^{(1)}}|I \rangle
+\langle I| \hat \mathcal{H}_{QI}|I \rangle
\nonumber
\\
& = & E_0^{pn} \ + \ E_{{MS}^{(1)}}^{[1]} + E_{QI}^{[1]}
.
\end{eqnarray}
Here, $E_{{MS}^{(1)}}^{[1]}$ is a correction to the monopole energy
$E_{0}^{pn}$ for a point nucleus due to ($s$- or
$p_{\frac{1}{2}}$-)electron penetration into the volume of a spherical
nucleus. The quadrupole interaction energy $E_{QI}^{[1]}$ is a
correction due to the deviation from spherical symmetry of this
nucleus.

There is a second group of entries with even much smaller corrections
in Tab.~\ref{corTab}: the HDI, QS$^{(1)}$ and MS$^{(2)}$ terms all
have $r^4$ and 4 derivatives of the electrostatic potential
($\rightarrow 1/r^{\,\prime 5}$). The corresponding Hamiltonians are:
\begin{widetext}
\begin{eqnarray}
\hat \mathcal{H}_{{HDI}} &=&
\frac{e H V_{zzzz}}{128I(I-1)(2I-1)(2I-3)\hbar^4}
\cdot \left[35\hat I_z^4 -30\hat I_z^2\hat I^2
+ 3\hat I^4 + 25\hbar^2I_z^2-6\hbar^2I^2
\right]
\label{HDIHami}
\\
\hat \mathcal{H}_{{QS}^{(1)}} &=&
-\frac{1}{14\epsilon_0} \frac{e\tilde Q n_{zz}}{4(2I-1)I\hbar^2}
\Big[\left(3\hat I_z^2 -\hat
  I^2\right)
\ \ + \frac{1}{2}\eta_{QS}\left(\hat I_+^2 + \hat I_-^2 \right)
  \Big]  \label{quadruSH}
\\
\hat \mathcal{H}_{{MS}^{(2)}} &=& -\frac{eZ}{5!\epsilon_0}
\Delta n(0) \
\langle r^4\rangle  \hat{1\!\mbox{l}}.
\label{MS2Hami}
\end{eqnarray}
\end{widetext}
The (diagonal part of the) hexadecapole Hamiltonian,
Eq.~(\ref{HDIHami}), is taken from the literature \cite{HDIHami}, the
quadrupole shift Hamiltonian, Eq.~(\ref{quadruSH}), is derived
explicitly in \cite{thesis} and similar algebra as for the first order
monopole shift Hamiltonian leads to the second order monopole shift
Hamiltonian, Eq.~(\ref{MS2Hami}).  As they are much smaller than the
QI and MS$^{(1)}$ terms, it makes little sense to add these
corrections to the Hamiltonian of Eq.~(\ref{eq-Hperturb}) right
away. Rather one should consider a first order perturbation to the
Hamiltonian of Eq.~(\ref{eq-Hperturb}), which itself was already a
perturbation to the monopole Hamiltonian of Eq.~(\ref{MIH}). This
means: find the perturbed eigen states of Eq.~(\ref{eq-Hperturb}) in
first order, and evaluate the new perturbations as given by the
Hamiltonians in Eqs.~(\ref{HDIHami})--(\ref{MS2Hami}) in these eigen
states. In the present work, we are interested in the first place in
$\hat{\mathcal{H}}_{{QS}^{(1)}}$, as it has the symmetry of a
quadrupole interaction: {\it this Hamiltonian, evaluated in the
  (approximate) eigenstates for a system with a finite and
  quadrupolarly deformed nucleus, gives an additional contribution to
  the regular quadrupole interaction.}
It can be interpreted as the influence on the quadrupole interaction
of electron penetration into the nuclear volume: the quadrupole
shift. The quadrupole shift Hamiltonian of Eq.~(\ref{quadruSH})
expresses the influence of the finite nucleus on the multipole
expansion. Evaluating this Hamiltonian for a density obtained from a
first-principles calculation with a finite nucleus expresses the
influence of the finite nucleus on the electronic wave functions.
%
%

There is an alternative way to express this same effect: consider the
Hamiltonian of Eq.~(\ref{eq-Hperturb}) up to second order
perturbation. Among others, the second order energy expression will
contain a cross term between QI and MS$^{(1)}$, which has the same
symmetry as the quadrupole interaction (this can be easily seen
because the Y$_{00}$ term of the monopole shift is a scalar quantity
that does not change the symmetry). Compared to the previous strategy
this method has the advantage that the same Hamiltonian is kept, but
the disadvantage that second order matrix elements in excited states
have to be evaluated. It is technically easier to evaluate a new
perturbation in the ground state of the previous perturbation. The
underlying physics, however, is the same.

The second order perturbation description has been applied in 1970 by
P.~Pyykk\"o for approximate and non-relativistic calculations in a few
test molecules (see also
Fig.~\ref{fig-logtrend})~\cite{Pyykko1970b}.
The first order + first order perturbation description has been used
in 2003 by Thyssen~{\it et al.}~\cite{Thyssen} for the case of LiI,
albeit in an implicit way that did not clearly showed the twofold
application of first order perturbation theory.  The twofold
application of perturbation theory will be the method used in the
present work as well, not at least because it leads to concise
analytical formulas. 
In 2006, also Karl and Novikov derived the so-called ``contact terms''
of the quadrupole interaction.  They used the Feynman diagram
technique and evaluated the results for hyperonic
atoms~\cite{K+N1,K+N2}.
Our derivation was made completely independent from the ones by
Thyssen~{\it et al.} and Karl and Novikov, and the observation that
the final expressions agree is a strong test of mutual correctness.
%

%
%
%
%
%
%
%
%
%
%
%

\subsection{Zooming in on $\mathbf{E_2}$}
\label{E2}

The regular quadrupole interaction and first order quadrupole shift
together provide our approximation to $E_2$:
\begin{eqnarray}
\nonumber
E_2  \ = \ h \nu_Q  \ \approx  \ E_{QI} \ + \ E_{QS^{(1)}}
 \ =  \ h \nu_{QI}+h \nu_{QS^{(1)}}.
\\
\label{nuQandnuQS}
\end{eqnarray}
Both terms consist each of a product between a nuclear quantity and an
electron quantity. As this shows to which nuclear and/or electronic
properties one gets access by measuring $E_2$, we discuss them
now. The two relevant nuclear quantities are (see
Eqs.~(\ref{eq-regular-QI-hamiltonian}) and~(\ref{quadruSH})):
\begin{eqnarray}
\hat{\mathcal{H}}_{QI} & \rightarrow & eQ \ = \ \int \rho (\vec r)
(3z^2-r^2) d \vec r \ \propto \ \langle r^2
Y_{20}\rangle
\nonumber
\\
\\
\hat{\mathcal{H}}_{{QS}^{(1)}} & \rightarrow &
e\tilde{Q} \ = \ \int \rho (\vec r) (3z^2-r^2){r^2} d \vec r \ \propto
\langle r^4 Y_{20}\rangle.
\nonumber
\\
\label{eq-nucl-QIMI}
\end{eqnarray}
The quasi quadrupole moment $\tilde Q$ has an additional $r^2$ in the
integral compared to the quadrupole moment $Q$. It is therefore a
quantity that bears similarity with the quadrupole moment $\langle r^2
Y_{20}\rangle$ (through the $Y_{20}$-dependence) as well as with the
hexadecapole moment $\langle r^4 Y_{40} \rangle$ (through the
$r^4$-dependence).

The corresponding electronic quantities are:
\begin{eqnarray}
\hat{\mathcal{H}}_{QI} & \rightarrow & V_{zz} \ =
\ \left(\partial_{zz} -
\frac{\Delta}{3}\right)v(0) \label{eq-Vzzoperator}
\\ \hat{\mathcal{H}}_{{QS}^{(1)}} & \rightarrow & n_{zz} \ =
\ {-\frac{1}{\epsilon_0}}\left(\partial_{zz}- \frac{\Delta}{3} \right)
  {\Delta} v(0). \label{eq-nzzoperator}
\end{eqnarray}
The integrated quantity (cf. Sec.~\ref{subsec-with-overlap}) $V_{zz}$
is the principal component of the electric-field gradient tensor. The
point quantity $n_{zz}$ is the main component of the tensor $n_{ij}
=(\partial_i\partial_j - \frac{\Delta}{3}\delta_{ij})n(0)$,
which has via the Laplacian two derivatives more than the main
component of the EFG tensor $V_{zz}$.  $n_{zz}$ can be shown to be
proportional to $\langle Y_{2m}/r^5 \rangle$ and therefore bears
similarities with the electric quadrupole field $\langle Y_{2m}/r^3
\rangle$ as well as with the electric hexadecapole field $\langle Y_{4m}/r^5
\rangle$, cf.~Eq.~(\ref{Vlm}).

%
%
%
%
%
%
%
%
%
%
%

\section{Observable consequences}
\label{subsec-overlap-consequences}
All entries in the (classical) Tab.~\ref{corTab} correspond to an
experimentally observable correction to the total energy. The first
row lists energy corrections which are a product of scalar
quantities. The leading term after the monopole contribution MI (or
$E_0^{pn}$) is the first order monopole shift MS$^{(1)}$, which
experimentally manifests its presence in the well-known isomer. The
second order monopole shift MS$^{(2)}$ is only very rarely taken into
account.  One example where it matters is the case of muonic
atoms~\cite{steffen1,steffen2} (atoms where a muon rather than an
electron orbits the nucleus). Because a muon is much heavier than an
electron, its orbit is much smaller and the overlap with the nuclear
charge distribution becomes much larger. This makes the second order
monopole shift for muons much larger than it is for
electrons~\footnote{To exploit the quadrupole shift (see
  Sec.~\ref{sec-exp-impl}) for muonic atoms, however, is not possible
  since no such experiments have been published for a couple of
  decades and the apparatus has been demounted~\cite{Pyykkoe2008}.}.

\begin{figure}[t]
\psfrag{Isub}{$I=\frac{5}{2}$ }
 \psfrag{m1sub}{$m_I=\pm\frac{5}{2}$ }
 \psfrag{m2sub}{$m_I=\pm\frac{3}{2}$ }
 \psfrag{m3sub}{$m_I=\pm\frac{1}{2}$ }
 \psfrag{n}{$\nu_{QI}$}
 \psfrag{2n}{$2\nu_{QI}$}
 \psfrag{nt}{$\nu_{QI}+\nu_{QS}$}
 \psfrag{2nt}{$2(\nu_{QI}+\nu_{QS})$}
\begin{center}
\includegraphics[width=\columnwidth]{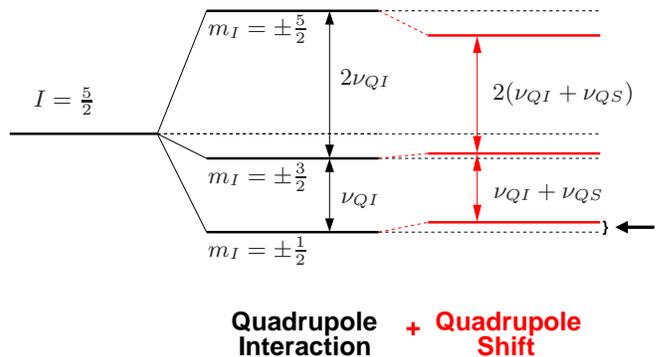}
\caption{Energy levels for a nuclear spin of $I=5/2$. This picture is
  not on scale: the shift of the levels as indicated by the arrow is
  in the most favorable cases (=heavy nuclei) only 0.1~\% of
  $\nu_{QI}$.}
\label{energylevel}
\end{center}
\end{figure}

All entries in the second row of Tab.~\ref{corTab} are dot products
between spherical tensors of rank 2. The first one is the quadrupole
interaction term QI, which splits according to
Eq.~(\ref{eq-regular-QI-hamiltonian}) energy levels that were
degenerate under the monopole term. An example for the axially
symmetric case ($\eta=0$) and nuclear spin $I=5/2$ is given in
Fig.~\ref{energylevel}. The second term in the second row is the first
order quadrupole shift $QS^{(1)}$, which shifts the energy levels that
were split by the quadrupole Hamiltonian, but preserves its overall
symmetry (Fig.~\ref{energylevel}, example for $\eta=\eta_{QS}=0$
\footnote{For all spherical tensors of rank two with a three, four or
  six fold rotation axis it can be shown, that only the $m=0$
  component is nonzero (which is equivalent to $\eta=0$)
  \cite{HFIbook}. This means, if $\eta=0$ then also $\eta_{QS}=0$ and
  vice versa.}).
The frequencies (energies) that set the scale for the quadrupole and
quadrupole shift splitting are (still considering $\eta=\eta_{QS}=0$):
\begin{eqnarray}
\label{nuQandnuQS-QI}
\nu_{QI} & = & \frac{eQV_{zz}}{h} \\
\label{nuQandnuQS-QS}
\nu_{QS} & = & -\frac{e\tilde Q n_{zz}}{14\epsilon_0 h}.
\end{eqnarray}
(For the sake of lighter notation, we will use from here on $\nu_{QS}$
rather than $\nu_{{QS}^{(1)}}$: we will not consider second order
quadrupole shifts and therefore no confusion will be possible.) The
quadrupole shift does not change the overall symmetry, which in the
example of Fig.~\ref{energylevel} means that the 1:2 ratio between the
two energy differences is preserved. An experiment that measures such
energy differences is not able to distinguish between the contribution
by $\nu_{QI}$ and the one by $\nu_{QS}$: it measures their sum only.
A discussion of the trends in the order of magnitude of the quadrupole
shift will be given in Sec.~\ref{subsec-trendsQS}, and several
experimental and computational strategies to exploit the quadrupole
shift will be suggested in Sec.~\ref{sec-exp-impl}.

Finally, the third row in Tab.~\ref{corTab} lists dot products between
tensors of rank 4. The leading term here is the hexadecapole
interaction for point nuclei. This term can in principle be
distinguished experimentally from a quadrupole interaction because its
symmetry is different (for instance, in Fig.~\ref{energylevel} the 1:2
ratio would be slightly violated). The HDI appears only for nuclei
with $l\geq2$, since only they have hexadecapole moments ($2I\geq l$
rule for $2^l$ multipole moments). Whereas the QI is well known and
experimentally accessible by e.g.\ NMR or Molecular Beam Spectroscopy
(see Sec.~\ref{AoQIE}), the situation for the HDI is different. Since
it was reported in 1955 for the first time \cite{HDI}, it has gone
through cycles of confirmatory measurements and refutations. An
overview is given in Ref.~\cite{Thyssen}.

%
%
%
%
%
%
%
%
%
%
%

\section{Computational aspects}
\label{sec-comp-aspects}

\subsection{Formulation in spherical notation}
\label{subsec-spherical}
The electronic part $n_{zz}$ of the quadrupole shift will be
calculated with a {first-principles} code and must therefore be
translated in spherical form as it is common in such codes:
\begin{eqnarray}
\label{nzz}
n_{zz}=\frac{2}{\sqrt{3}}\sqrt{\frac{15}{4\pi}}
\lim_{r\rightarrow 0}\frac{1}{r^2}n_{20}(r).
\end{eqnarray}
The spherical component of the density, $n_{20}(r)$, which enters this
expression, is the radial part of the $\left(l\!=\!2,\,m\!=\!0\right)$
component of expansion of the density $n(\vec r)$ in spherical
harmonics:
\begin{eqnarray}
n(\vec r)=\sum_{lm}n_{lm}(r)Y_{lm}(\Omega).
\end{eqnarray}
The $l\!=\!2$ components are closely related to Cartesian second
derivatives~\cite{Koch}, which is the reason why they appear in the
electric-field gradient and related quantities.

%
%
%
%
%
%
%
%
%
%
%
\subsection{Computational details: the FPLO code}  
\label{FPLOmeth}

All calculations in this paper have been performed by the Density
Functional Theory solid state code FPLO~\cite{FPLO} (version 8.00-31),
which is is a full-potential band structure scheme and based on linear
combinations of overlapping non-orthogonal atom-centred orbitals. The
core relaxation is properly taken into account (so called all-electron
method).  We used the Local Density Approximation (LDA) for the
exchange-correlation functional~\cite{xc-PW}.  FPLO can perform
non-relativistic, scalar-relativistic as well as fully-relativistic
calculations~\cite{FPLOdocu2,FREL}. In the latter, the Dirac
Hamiltonian with a general potential is solved. Recently, a finite
nucleus has been implemented in FPLO, which is crucial for the
present work (Sec.~\ref{subsec-finuc}).

%
%
%
%
%
%
%
%
%
%

\subsection{Relativity and the role of a finite nucleus}
\label{subsec-finuc}
\begin{figure}[t]
\begin{center}
\includegraphics[width=\columnwidth]{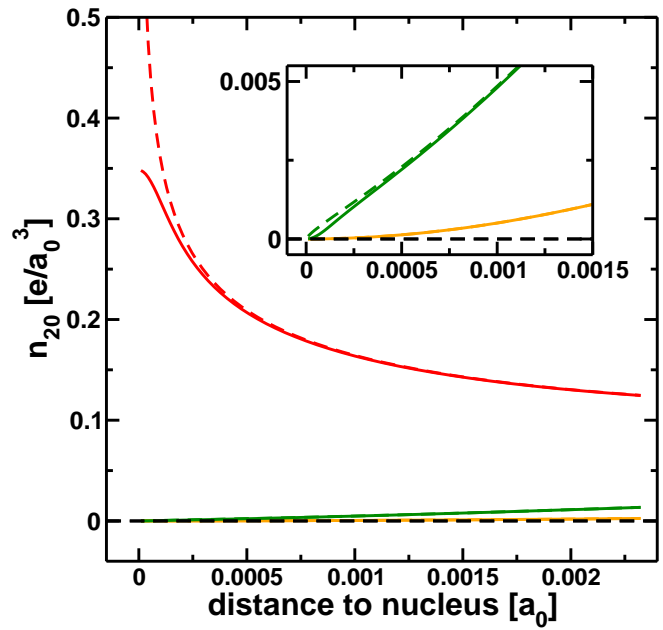}
\end{center}
\caption{The density component $n_{2m}(r)$ including inset, which
  zooms in the region around $r=0$ ) for a point nucleus (p.n., dashed
  lines) and a finite nucleus (f.n., full lines) plotted in dependence
  of $r$.  The different methods are indicated by different colors:
  non-relativistic (NREL, yellow), scalar relativistic (SREL, red) and
  full relativistic (FREL, green). This calculation for the hcp metal
  Re was done by FPLO. All quantities are given in atomic units.}
\label{Nucpic1}
\end{figure}

In order to obtain $n_{zz}$, the limit of $n_{20}(r)/r^2$ for
$r\!\rightarrow\!0$ must be calculated, cf.~Eq.~(\ref{nzz}).  It
matters whether this is done within a non-relativistic (NREL), a
scalar-relativistic (SREL) or a fully relativistic (FREL) framework.%
%
%
In the NREL or FREL formulations (no matter if a point or a finite
nucleus is used in the calculation) $n_{2m}(0)$ is exactly zero as it
should be due to angular selection rules (Fig.~\ref{Nucpic1}).
In the SREL approximation, the $\left(l\!=\!2,m\right)$ density,
created from two divergent $p_{1/2}$ functions, is to some extent
wrongly non-zero at $r\!=\!0$.  This makes SREL-based methods (with or
without a point nucleus) essentially useless for calculating
properties that depend on $n_{20} (r\!\rightarrow\!0)$, and we will
therefore not consider SREL any further.

\begin{figure}[t]
\begin{center}
\includegraphics[width=\columnwidth]{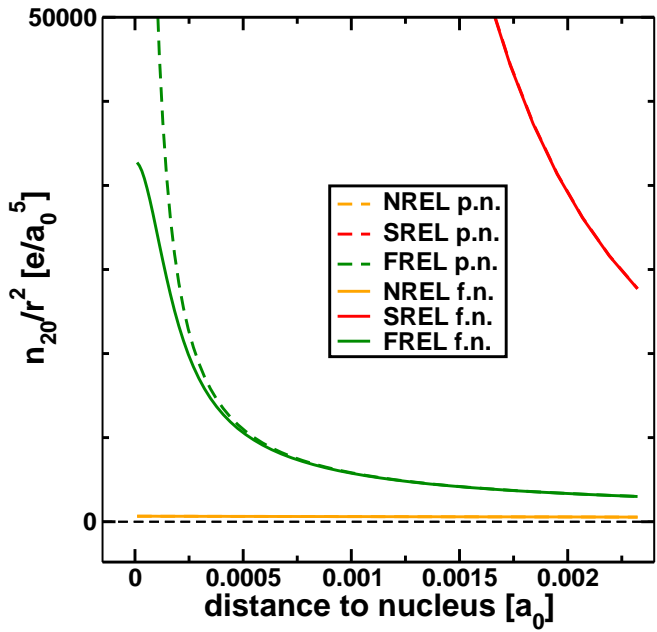}
\end{center}
\caption{The density component $n_{2m}(r)/r^2$ for a point nucleus
  (p.n., dashed lines) and a finite nucleus (f.n., full lines) plotted
  in dependence of $r$.  The different methods are indicated by
  different colors: non-relativistic (NREL, yellow), scalar
  relativistic (SREL, red) and full relativistic (FREL, green). This
  calculation for the hcp metal Re was done by FPLO. All quantities
  are given in atomic units.}
\label{Nucpic2}
\end{figure}

For a point nucleus, the ratio of $n_{20}(r)$ and $r^2$ converges for
the limit $r\!\rightarrow\!0$ in a NREL formulation, but not in a FREL
formulation (Fig.~\ref{Nucpic2}).  Since this ratio
at $r\!=\!0$ is an observable quantity (see Eqs.~(\ref{nzz})
and~(\ref{nuQandnuQS-QS})), the divergence for the better method (FREL
vs.\ NREL) cannot be physical. And indeed, the divergence disappears
if the approximation of a point nucleus is dropped and a finite
nucleus is used in the calculation (Fig.~\ref{Nucpic2}).
Numerical values for this ratio turn out to be much larger for FREL
compared to NREL, especially for heavy elements.

The divergence of $n_{zz}$ in a fully relativistic point nucleus
calculation might appear to be worrying at first sight. Wouldn't that
mean that the quadrupole shift in Eq.~(\ref{nuQandnuQS-QS}) is
infinite? The answer is: no, because the operator corresponding to
$n_{zz}$ (Eq.~(\ref{eq-nzzoperator})) does not have to be evaluated in
the ground state for the point nucleus (which is the case that
diverges at $r=0$), but in the ground state after having added the two
perturbations of Eq.~(\ref{eq-Hperturb}) that describe the effect of
quadrupolarly deformed finite nucleus (where the divergence is
absent). The latter ground state can be constructed from the ground
and excited states of the point nucleus case, applying the common
expression for the eigenfunctions in first order perturbation. This
would, however, lead to rather lengthy expressions and to the
inconvenience of having to use excited states. A pragmatic workaround
is to use instead the ground state as calculated in a
{first-principles} code that takes a finite nucleus into account. This
is hardly an approximation, as it was exactly the purpose of the
perturbations in Eq.~(\ref{eq-Hperturb}) to express the presence of a
finite nucleus. Therefore, we conclude that {\it the quadrupole shift
  can be obtained by evaluating the operator for $n_{zz}$ in
  Eq.~(\ref{eq-nzzoperator}) for the ground state of the atom,
  molecule or solid calculated fully relativistically and with a
  finite nucleus taken into account.} This quadrupole shift has to be
added to the contribution obtained by evaluating the operator for
$V_{zz}$ in Eq.~(\ref{eq-Vzzoperator}) in the ground state of the
point nucleus case (and not in the ground state of the finite nucleus
case, as the regular QI is really a perturbation to the point
nucleus).

%
%
%
%
%
%
%
%
%
%
%

\subsection{Comparison with the PCNQM method}
\label{subsec-PCNQM}
We have described in the previous sections a procedure to obtain the
influence on the quadrupole interaction of electron penetration in a
finite nucleus by two subsequent applications of first order
perturbation theory combined with finite nucleus calculations
(Eq.~(\ref{nuQandnuQS-QS}) and Figs.~\ref{Nucpic1} and
~\ref{Nucpic2}). An alternative to this procedure is the {\it point
  charge nuclear quadrupole moment} method
(PCNQM)~\cite{Pernpointner1998,Kello1998c,Kello2000b}: the
electric-field gradient is not obtained as the expectation value of an
operator, but is determined from the way how the total energy of the
system changes upon inserting an artificial array of point charges
around the nucleus. In this method, only total energies are required
to obtain the electric-field gradient, which makes it particularly
useful when the proper operator for the field gradient is not
explicitly known. The latter is for instance the case as soon as a
finite nucleus is used (Eq.~(\ref{eq-regular-QI-hamiltonian}) is valid
for a point nucleus only), or for fully relativistic calculations at
the 2-component level (a complicated and not yet performed `picture
change' transformation would be needed to find the 2-component version
of the EFG operator~\cite{Pernpointner1998}.) The difference in EFGs
between a `finite nucleus + PCNQM' calculation and a point nucleus
calculation (either with the regular EFG operator or with PCNQM) gives
the effect of electron penetration in the nucleus. One case where this
difference is explicitly calculated is for $^{127}$I in LiI
(Ref.~\cite{VanStralen2003} and Fig.~\ref{fig-logtrend}). With
PCNQM, the quadrupole shift can be obtained only numerically: there is
no analytical expression as Eq.~(\ref{nuQandnuQS-QS}).

In passing, we note here that a method that is quite analogous to
PCNQM has been recently developed for the first order monopole shift
correction MS$^{(1)}$ (isotope shift, isomer
shift)~\cite{Filatov2007,Filatov2008} as well.

%
%
%
%
%
%
%
%
%
%
%
\section{Numbers and trends}
\label{sec-implementation-trends}

In the present section, we will perform actual calculations with the
formalism described in Secs.~\ref{sec-formalism}
and~\ref{sec-comp-aspects}, and examine trends in the relevant
quantities: the nuclear quasi-quadrupole moment $\tilde{Q}$, the
electronic point property $n_{zz}$, and their product: the quadrupole
shift $\nu_{QS}$.

%
%
%
%
%
%
%
%
%
%
%

\subsection{Trends in $\mathbf{\tilde{Q}}$}\label{trendsintildeQ}

Consider a phenomenological model for a nucleus: a deformed sphere,
with a radius $R(\theta)$ given by ~\cite{Ring1980}:
\begin{eqnarray}
\label{eq-nuclear-shape}
R(\theta)=a \left( 1+\beta_2 Y_{20}(\theta)+\beta_4Y_{40}(\theta) +
\dots \right),
\end{eqnarray}
where $a$ is called the monopole radius and the $\beta_i$ are
deformation parameters. The monopole radius depends in the first place
on the atomic mass number $A$ of the nucleus, and the main trend
through a lot of experimental values can be summarized
by~\cite{Agneli2004}\footnote{Note, that in Ref.~\cite{Agneli2004},
  the root-mean-square (RMS) of the nuclear radius is given, while
  here we use the monopole radius.}
\begin{eqnarray}
\label{nucradius}
a(A)=  1.489 \ A^{0.294}~\mbox{fm}.
\end{eqnarray}
Values for $\beta_2$ fall rarely outside the range $[-0.3,\,
  +0.3]$ (Ref.~\cite{Stone2005} in combination with
Eq.~(\ref{Qbeta_2})). As $\beta_4$ is even smaller and enters only
quadratically in the expressions we will need, it can be neglected for
our purposes. Keeping only the linear order for $\beta_2$, we can now
express the quadrupole moment and the quasi quadrupole moment in terms
of $a$ and $\beta_2$:
\begin{eqnarray}
\label{Qbeta_2}
e Q & \simeq & 3\sqrt{\frac{4\pi}{5}}\frac{eZ}{2\pi}\beta_2 a^2 \\
e\tilde Q & \simeq & a^2\cdot eQ \label{eq-tildeQ}.
\end{eqnarray}
The term quadratic in $\beta_2$ as well as the quadratic $\beta_4$
term give corrections to Eqs.~(\ref{Qbeta_2}) and~(\ref{eq-tildeQ}) at
the level of a few percent only, while they make the expressions
considerably more involved -- see Ref.~\cite{thesis}.

By Eqs.~(\ref{Qbeta_2})-(\ref{eq-tildeQ}), one can get a reasonable
estimate for $\tilde Q$ by inserting the monopole radius from
Eq.~(\ref{nucradius}) and the experimental quadrupole moment $Q$
(e.g.\ from Ref.~\cite{Harris2002,Stone2005,Pyykkoe2008}). In this
way, we obtain values for $\tilde Q$ in the order of
$10^4-10^5$~fm$^4$ for heavy elements (Tab.~\ref{Qtildetab}).
\begin{table}[t]
\caption{The nuclear radius $a$, quadrupole moment $Q$, deformation
  parameter $\beta_2$ and quasi quadrupole moment $\tilde Q$ of a few
  isotopes.}
\label{Qtildetab}
\begin{center}
 \begin{tabular*}{\linewidth}{r@{\extracolsep\fill}rrrr}
\hline
\hline
\\[-0.25cm]
 Isotope  & $a$ [fm] &  $Q$ [fm$^2$] & $\beta_2$  &  $\tilde Q$ [fm$^4$]
\\[.15cm]
\hline
\\[-0.35cm]
$^{9}$Be  & 2.84 & 5.3  & 0.22  & 43
\\[.05cm]
$^{47}$Ti   & 4.61  & 30.2  & 0.09 &  644
\\[.05cm]
$^{111}$Cd   & 5.95 & 83.0  & 0.07 & 2$\ \!$934
\\[.05cm]
$^{138}$La  & 6.34 &  45.0    &   0.03 & 1$\ \!$808
\\[.05cm]
$^{179}$Hf  & 6.84 & 379.3   &   0.15 & 17$\ \!$760
\\[.05cm]
$^{187}$Re  & 6.93 & 207.0   &   0.07 & 9$\ \!$945
\\[.05cm]
$^{189}$Os  & 6.95 &  85.6   &   0.03 & 4$\ \!$138
\\[.05cm]
\hline
\hline
\end{tabular*}
\end{center}
\end{table}
The Eqs.~(\ref{Qbeta_2})-(\ref{eq-tildeQ}) show that in order to get a
large quasi quadrupole moment~$\tilde{Q}$, the nucleus should be large
($a$ is large) and strongly deformed ($Q$ or $\beta_2$ are large).
The former implies heavy elements, while the latter is most easily
fulfilled for heavy elements as well.

\subsection{Trends in $\mathbf{n_{zz}}$}
\label{trendsinnzz}

In order to get a feeling for the order of the magnitude of the
electronic parts of the $\mathcal{O}(2)$ interactions in
Tab.~\ref{corTab}, we have calculated both $V_{zz}$ (the electronic
part of the QI) and $n_{zz}$ (the electronic part of the first order
QS) for some hexagonal close-packed (hcp) metals throughout the
periodic table. The results are shown in Tab.~\ref{QandQStable}. Both
quantities increase with the mass of the element. But compared to
$V_{zz}$, which increases over two orders of magnitude, $n_{zz}$ is
much more sensitive to the mass of the element and increases over
eight orders of magnitude.

In order to verify to which extent this conclusion obtained from
Tab.~\ref{QandQStable} is valid for other crystal structures than hcp,
we investigated two series of purpose-built body-centered tetragonal
(bct) crystals with different $c/a$ ratios (0.8 and 1.2), and this for
several elements throughout the periodic table. The results are
reported in Ref.~\cite{thesis} and show the same trend as
Tab.~\ref{QandQStable}. We conclude that the mass of the element has a
larger influence on the magnitude of $n_{zz}$ than the lattice
parameters or the crystal structure.

%
%
%
%
%
%
%
%
%
%
%

\subsection{Trends in the quadrupole shift}
\label{subsec-trendsQS}

\begin{figure}[t]
\begin{center}
\includegraphics[width=\columnwidth]{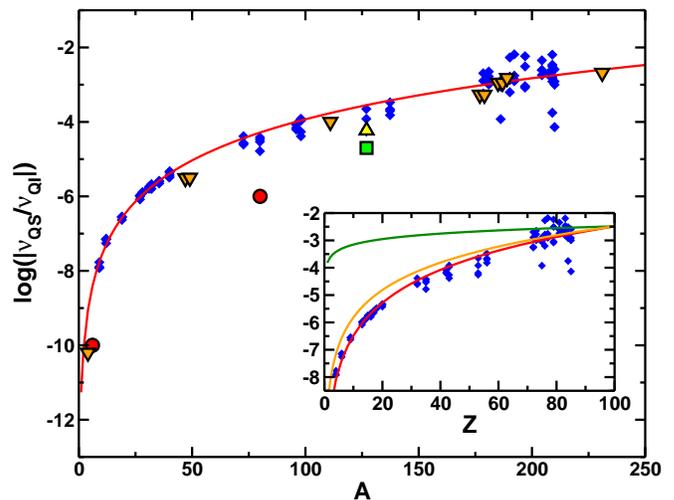}
\caption{The logarithm of the ratio of $\nu_{QS}$ and $\nu_{QI}$ as a
  function of the mass number $A$. Blue diamonds: artificial crystal
  structures (see text and Ref.~\cite{thesis}), fitted by the red line
  (Eq.~(\ref{eq-A4})). Orange triangles (down): experimental crystal
  structures (see Tab.~\ref{QandQStable}). The yellow triangle
  (up)~\cite{Thyssen}, green square~\cite{VanStralen2003} and red
  circles~\cite{Pyykko1970b} are values from the literature, see
  text. {\bf Inset:} the same data but now as a function of $Z$, fit
  by Eq.~(\ref{eq-Z4}). The nuclear (green) and electronic (orange)
  contributions of Eq.~(\ref{eq-Z4analytic}) are shown as well,
  shifted to match in the endpoint.}
\label{fig-logtrend}
\end{center}
\end{figure}
\begin{table*}[t]
\caption{For a few atoms/nuclei that experimentally condense in the
  hcp crystal structure (except for Pa, bct), this table lists the
  nuclear properties $Q$ and $\tilde Q$ (determined as in
  Table~\ref{Qtildetab}), the electronic properties $V_{zz}$ and
  $n_{zz}/\epsilon_0$ (calculated by FPLO, see text), the quadrupole
  $\nu_{QI}$ and quadrupole shift $\nu_{QS}$ frequencies they give
  rise to (Eqs.~(\ref{nuQandnuQS-QI}) and~(\ref{nuQandnuQS-QS})) (mind
  the different MHz and kHz units), and the ratio of the latter.}
\label{QandQStable}
\begin{center}
\begin{tabular*}{\linewidth}{r@{\extracolsep\fill}rrrrrrrrrr}
\hline
\hline
\\[-0.25cm]
   Isotope &   $I$ &  $Q$ & $\tilde Q$ & $V_{zz}$ & $n_{zz}/\epsilon_0$
& $\nu_{QI}$ & $\nu_{QS}$  &   $\left|{\nu_{QS}}/{\nu_{QI}}\right|$
\\
 &  &\tiny [fm$^2$] & \tiny[fm$^4$] & \tiny [$10^{21}$V/m$^2$]
&\tiny [$10^{42}$V/m$^4$]
 & \tiny [MHz]&  \tiny[kHz] &
\\[.15cm]
\hline
\\[-0.35cm]
$\!\!\!^{9}$Be   & \ ${3}/{2}$   & 5 &   42 &{ $-$0.08}
& $-$6.07$\cdot 10^{-2}$& $-$0.1 & $10^{-8}$ &   5$\cdot 10^{-9}$
\\[.05cm]
$\!\!\!^{47}$Ti   &  ${5}/{2}$   & 30 &  644 & 1.61
&3.27$\cdot 10^{+3}$&11.8 & $-$0.04 &   3$\cdot 10^{-6}$
\\[.05cm]
$\!\!\!^{49}$Ti   & ${5}/{2}$   & 25 &  539 & 1.61
&3.27$\cdot 10^{+3}$&9.6 & $-$0.04 &   3$\cdot 10^{-6}$
\\[.05cm]
$\!\!\!^{111}$Cd   & ${5}/{2}$   & 83 &    2$\ \!$934 & 7.48
&2.94$\cdot 10^{+5}$&150.0 & $-$14.9 &   1$\cdot 10^{-4}$
\\
[.05cm]
$\!\!\!^{177}$Hf   & ${7}/{2}$  & 337 &   15$\ \!$652 & 7.89
&1.26$\cdot 10^{+6}$&642.3 & $-$341.4 &   5$\cdot 10^{-4}$
\\ [.05cm]
$\!\!\!^{179}$Hf   & ${9}/{2}$   & 379 &    17$\ \!$760 & 7.89
&1.26$\cdot 10^{+6}$&723.9 & $-$387.4 &   5$\cdot 10^{-4}$
\\ [.05cm]
$\!\!\!^{185}$Re   & ${5}/{2}$   & 218 &   10$\ \!$386 & $-$5.51
& $-$1.81$\cdot 10^{+6}$& $-$290.3 & 324.9  &  1$\cdot 10^{-3}$
\\ [.05cm]
$\!\!\!^{187}$Re   & ${5}/{2}$   & 207 &    9$\ \!$945 & $-$5.51
& $-$1.81$\cdot 10^{+6}$& $-$275.6 & 311.1  &  1$\cdot 10^{-3}$
\\ [.05cm]
$\!\!\!^{189}$Os   & ${3}/{2}$   &  86 &    4$\ \!$138 & $-$6.65
& $-$2.91$\cdot 10^{+6}$& $-$137.6 & 208.1 &  2$\cdot 10^{-3}$
\\ [.05cm]
$\!\!\!^{231}$Pa   & ${3}/{2}$   &  $-$172 &   -9$\ \!$357 &15.14
& 8.11$\cdot 10^{+6}$& $-$629.8 & 1309.7 &  2$\cdot 10^{-3}$
\\[.05cm]
\hline
\hline
\end{tabular*}
\end{center}
\end{table*}
How do the nuclear part from Sec.~\ref{trendsintildeQ} and the
electronic part from Sec.~\ref{trendsinnzz} combine to produce a
quadrupole shift? The frequencies $\nu_{QI}$ (for the QI --
Eq.~(\ref{nuQandnuQS-QI})) and $\nu_{QS}$ (for the QS --
Eq.~(\ref{nuQandnuQS-QS})) for a set of hcp and bct metals are
reported in Tab.~\ref{QandQStable}, together with their ratio
$\left|{\nu_{QS}}/{\nu_{QI}}\right|$. Experimental lattice parameters
were used~\cite{hcpKKR,Pa}, and $n_{zz}$ and $V_{zz}$ were determined
fully relativistically with a finite nucleus for $n_{zz}$ and a point
nucleus for $V_{zz}$ (see Sec.~\ref{FPLOmeth}). $Q$ was taken from the
literature \cite{Pyykkoe2008} and $\tilde Q$ was determined as
explained in Sec.~\ref{trendsintildeQ}.
The trends of $n_{zz}$ and $\tilde{Q}$ to be larger for heavy
elements, cooperate to produce a $\nu_{QS}$ of which the relative
importance with respect to $\nu_{QI}$ is rather smoothly increasing
with the atomic number $A$.

This can be seen more clearly in Fig.~\ref{fig-logtrend} (blue
diamonds), which summarizes results for a larger set of 28 elements in
different crystal structures: hcp with c/a=1.633 and 0.8 and bct with
c/a=1.2 and 0.8, always with the experimental volume per atom (details
are given in Ref.~\cite{thesis}). These data can be fit with the
simple functions
\begin{eqnarray}
\label{eq-A4}
|\nu_{QS}| & = & 5.46 \cdot 10^{-12}  \, A^{\frac{11}{3}} \, |\nu_{QI}| ,
\\
|\nu_{QS}| & = & 3.26 \cdot 10^{-11}  \, Z^4 \, |\nu_{QI}| ,
\label{eq-Z4}
\end{eqnarray}
which are shown in Fig.~\ref{fig-logtrend} (red lines). The orange
triangles (down) in Fig.~\ref{fig-logtrend} correspond to the
experimental crystal structures from Tab.~\ref{QandQStable} -- they
accurately follow the same trend.

By taking the ratio of Eqs.~(\ref{nuQandnuQS-QS})
and~(\ref{nuQandnuQS-QI}) and by filling out the lowest order
expressions for $Q$ and $\tilde{Q}$ (Eq.~(\ref{Qbeta_2})), the
following simple analytic analogue for Eqs.~(\ref{eq-A4})
or~(\ref{eq-Z4}) is obtained:
\begin{eqnarray}
\nu_{QS} & = & \left( - \frac{1}{14} \, a^2 \, \frac{n_{zz}}{\epsilon_0}
\, \frac{1}{V_{zz}} \right) \, \nu_{QI}.
\label{eq-Z4analytic}
\end{eqnarray}
Since $a=1.26\,Z^{1/3}$~fm (obtained from the data of
Ref.~\cite{Agneli2004} plotted as a function of $Z$), the nuclear part
$a^2$ scales with $Z^{2/3}$. In order to fulfill the observed $Z^4$
dependence in Eq.~(\ref{eq-Z4}), the electronic part should scale
\mbox{with $Z^{10/3}$: $n_{zz}/(\epsilon_0 V_{zz})= 2.87 \cdot
  10^{-10}\,Z^{10/3}$~fm$^{-2}$.} These two contributions are shown as
the green (nuclear) and orange (electronic) lines in the inset of
Fig.~\ref{fig-logtrend}. From this picture, it is clear that the
electronic term contributes most to the increase of the quadrupole
shift with $A$ or $Z$. From Tab.~\ref{QandQStable}, we see that this
is due to the strong increase of $n_{zz}$.

Eq.~(\ref{eq-Z4}) provides a quick way to estimate the order of
magnitude of the quadrupole shift, for any element in any crystal
structure, and without the need for a finite nucleus calculation. The
only quantity that is required is $\nu_{QI}$, which can be provided by
several {first-principles} codes. As the scatter of the data points
for heavier elements shows, such an estimate can be one order of
magnitude above or below the actual value. For isotopes with $A\gtrsim
175$ ($Z\gtrsim 60$), the quadrupole shift can reach 0.1-1.0\% of the
regular quadrupole interaction.
%

There are a few cases reported in the literature from which QS
information can be deduced. These are shown in Fig.~\ref{fig-logtrend}
as well. The yellow triangle (up) was calculated by J.~Thyssen~{\it et
  al.} with a method very similar to ours for the single case of the
LiI molecule. They found the ratio
$\left|{\nu_{QS}}/{\nu_{QI}}\right|$ for $^{127}$I to be
$5\cdot10^{-5}$. The green square corresponds to $^{127}$I in the same
LiI molecule, obtained by the PCNQM method by Van~Stralen and
Visscher~\cite{VanStralen2003}. A few estimates for the quadrupole
shift obtained by second order perturbation theory were published in
1970 by P.~Pyykk\"o~\cite{Pyykko1970b}. Those estimates were given
relative to a pseudo quadrupole interaction only
(\ref{subsec-pseudoQI} and Ref.~\cite{Pyykko1970}). After converting
these numbers, it turns out that for the LiBr molecule the ratio of
$\nu_{QS}$ and $\nu_{QI}$ is about 10$^{-10}$ for $^6$Li and 10$^{-6}$
for $^{81}$Br (red circles in Fig.~\ref{fig-logtrend}). These numbers
follow the same trend as the quadrupole shift in first order
perturbation, but are 1-2 orders of magnitude smaller -- this might be
due to the fact that these were non-relativistic calculations.

%
%
%
%
%
%
%
%
%
%
%

\subsection{Other small perturbations to the quadrupole interaction}
\label{sec-other}

When dealing with a quadrupole-like interaction that is as small as
the quadrupole shift, it becomes relevant to take into account
similarly small quadrupole-like interactions and perturbations of the
quadrupole interaction that have a different origin. Some of these
interactions have been proposed decades ago, in an era during which it
was impossible to compute accurate values for them. Given the enormous
advances in the possibilities of {first-principles} calculations
since that time, it is worthwhile to list these effects here, to
discuss them shortly, to put them into a general picture and to refer
to the original literature. This is done in App.~\ref{appendix}, where we
will deal with second order effects of magnetic origin, the
isotopologue anomaly and the influence of temperature.

%
%
%
%
%
%
%

\section{Experimental implications of the quadrupole shift}
\label{sec-exp-impl}
\subsection{Accuracy of quadrupole interaction experiments and calculations}
\label{AoQIE}

In order to see whether or not the presence of the quadrupole shift
can be experimentally detected, we should assess the accuracy that can
be achieved in quadrupole interaction experiments. In order to find
out which kind of information can be extracted if such experiments are
combined with {first-principles} calculations, the best achievable
accuracy in such calculations will be discussed as well.

Experimental methods can be either non-radioactive ones as nuclear
magnetic resonance (NMR) spectroscopy, nuclear quadrupole resonance
(NQR) spectroscopy, laser spectroscopy (LS) and molecular beam
spectroscopy (MBS), or radioactive ones as M\"ossbauer spectroscopy
and perturbed angular correlation (PAC) spectroscopy. These methods
can be applied to atoms and molecules (LS, MBS, NMR, NQR) or to solids
(NMR, NQR, M\"ossbauer, PAC). A typical NQCC $\nu_Q$ is of the order
of magnitude of 100~MHz. These are the lowest achievable experimental
error bars on $\nu_Q$ for each method: 5~kHz for NMR or NQR on single
crystals with an axially symmetric EFG \cite{LBnqr,NMRbook}, 100~kHz for NMR
or NQR on powder samples with a non-axially symmetric EFG
\cite{LBnqr,NMRbook}, 5~MHz for LS on atomic beams~\cite{Basar2009}, 5-20~Hz
(!)  for MBS~\cite{Cederberg1999,Cederberg2006a}, 500~kHz for
PAC~\cite{CdT1,CdT2,CdT3} and M\"ossbauer spectroscopy
\cite{accMos}. When compared to the the quadrupole shift values in
Tab.~\ref{QandQStable} (typically 100~kHz for A$\gtrsim$150), it is
clear that only NMR in solids and especially Molecular Beam
Spectroscopy on molecules are sensitive to the quadrupole shift --
provided the isotope under consideration is sufficiently heavy.

{First-principles} calculations in solids are commonly done at the level
of density functional theory (DFT), or with DFT as a starting
point. DFT has been used with considerable success to calculate
electric-field gradients in solids, see e.g.\ Refs.~\cite{EFG1985,
  EFG1988, Akai1990,Svane1997, EFGmotive1, PetrilliBlaha,EFGmotive3,
  EFGmotive2, EFGsuccess1, EFGsuccess2, hcpKKR, MGa2paper, 122paper,
  Ndpaper}. As a rule of thumb, the DFT prediction is within
10\% of the experimental value.

{First-principles} calculations for (small) molecules can resort
to Hartree-Fock calculations with correlation corrections. These are
computationally much more demanding, but can in principle provide an
arbitrary high precision. The recent literature \cite{Kelloe1998a,
  Kelloe1998b, Kelloe1999a, Kelloe1999b, Kelloe2000a, Pernpointer2000}
shows that correlation corrections using coupled cluster theory with
single, double and (perturbatively) triple excitations (CCSD(T)),
combined with sufficiently large basis sets and -- where needed --
with a \mbox{(semi-)}relativistic Hamiltonian, provides highly
accurate EFGs for small molecules.  It has been claimed
\cite{Kello2007} that in this way an absolute accuracy with 4
significant digits can be reached. This is considerably better than
the accuracy which DFT can provide for the EFG in solids. For
molecules that are too demanding for a CCSD(T) treatment, DFT with the
recently proposed CAMB3LYP* functional can be an
alternative~\cite{Thierfelder2007}. DFT for EFGs in small molecules
can be very unreliable~\cite{Thierfelder2007}.

At non-zero temperatures, vibrational states will be populated in
solids and molecules, and in molecules rotational states as well. This
will influence the electric-field gradient. In solid state
calculations, this has so far only rarely been taken into
account~\cite{Torumba2006}. In molecules, the effect of temperature is
routinely taken into account in
calculations~\cite{Kelloe1999b,Schwerdtfeger2005} as well as in the
analysis of
experiments~\cite{Cederberg2005,Schlier1961,Dunham1932}. This allows
an even more detailed comparison between experiment and theory for
molecules.

%
%
%
%
%
%
%
%
%

\subsection{Determination of $\mathbf{Q}$ and $\mathbf{\tilde{Q}}$: method}
\label{subsec-QtQ-method}
With the experimental accuracies listed in the previous section, it is
clear that experimental nuclear quadrupole coupling constants $\nu_Q$
for NMR on single crystals and for molecular beam spectroscopy on
molecules are affected by the quadrupole shift. This means that the
experimentally determined value for $\nu_Q$ would have a different
value (outside the error bar) if the quadrupole shift could be
``switched off". It does not mean, however, that by such an experiment
the quadrupole shift itself can be determined: the QS manifests itself
as an addition to the regular quadrupole interaction, and is
indistinguishable from it (Eqs.~(\ref{nuQandnuQS-QI})
and~(\ref{nuQandnuQS-QS})):
\begin{eqnarray}
\nu_{Q}& \approx & \nu_{QI} \, + \, \nu_{QS} \ = \ Q \, \frac{eV_{
    {zz}}}{h} - \tilde{Q} \frac{e n_{ {zz}}}{14 \epsilon_0 h} .
\label{eq-linear2}
\end{eqnarray}
The second term of this equation is even in the most favorable cases
2-3 orders of magnitude smaller than the first term
(Fig.~\ref{fig-logtrend}). If $\nu_Q$ is measured, if $V_{zz}$ is
calculated from {first-principles} and if $\nu_{QS}$ is neglected,
then the quadrupole moment $Q$ can be determined from
Eq.~({\ref{eq-linear2}). This has become the preferred procedure to
  determine nuclear quadrupole moments
  (e.g.~\cite{Dufek1995,Svane1997,Pyykkoe2008,
    Kelloe2000c,Thierfelder2007,Pyykkoe2008,QMofTc,Barone2009}).

If $V_{zz}$
  could be calculated with an arbitrary high precision, the precision
  of the resulting $Q$ is limited by neglecting $\nu_{QS}$. One could
  choose not to neglect $\nu_{QS}$, and apply Eq.~(\ref{eq-linear2})
  to at least two $\nu_Q$ measurements in order to determine
  simultaneously a more precise value of $Q$ and $\tilde{Q}$ (or $Q$
  and $a^2$). This would be meaningful only in cases where the
  absolute deviations on the computed $V_{zz}$ and $n_{zz}$ values are
  small enough to make the uncertainty in $\nu_{QI}$ smaller than the
  value of $\nu_{QS}$. The only hope to realize this is in the case of
  sufficiently heavy elements, for which, however, it might not yet be
  feasible to achieve the requested computational accuracy.

\subsection{Quadrupole moment ratios: the quadrupole anomaly}
When it is not possible to know experimentally the value of a
quadrupole moment with sufficient accuracy, the next best thing to
know are ratios of quadrupole moments for two different isotopes, or
for two different isomeric states of the same isotope. As soon as a
later experiment succeeds to determine one of the quadrupole moments
in the ratio, the other one is known as well.

The ratio $Q_1/Q_2$ of two quadrupole moments is commonly measured as
the ratio $\nu_{Q,1}/\nu_{Q,2}$ of two nuclear quadrupole coupling
constants.  Indeed, in the absence of a quadrupole shift, both ratios are identical
if the two isotopes or isomers are in the same environment and
therefore experience the same $V_{ {zz}}$
(Eq.~({\ref{eq-linear2})). The presence of the quadrupole shift,
  however, spoils the equality of both ratios. It is straightforward to show that the
  ratio of quadrupole coupling constants is equal to
\begin{eqnarray}
\frac{\nu_{Q,1}}{\nu_{Q,2}} & = & \frac{Q_1}{Q_2} \left( 1 + \delta \right)
\nonumber
\\
 \mbox{with} \ \ \ \delta & = & \frac{n_{ {zz}}}{14
\epsilon_0 V_{ {zz}}} \, \left( a_2^2 - a_1^2 \right) +
O\left(a_i^4\right).
\label{eq-ratio}
\end{eqnarray}
This formulation is strongly reminiscent to the Bohr-Weisskopf effect
\cite{BohrWeisskopfEffect} for magnetic hyperfine interactions, where
the ratio between two {\it magnetic} hyperfine interaction frequencies
for two isotopes/isomers at identical sites is given by
\begin{eqnarray}
\label{deltaQS}
\frac{\nu_1}{\nu_2} & = & \frac{\mu_1}{\mu_2} \,
\left( 1 + \Delta \right).
\end{eqnarray}
Here $\mu_1$ and $\mu_2$ are the nuclear magnetic moments of the two
isotopes/isomers, and $\Delta$ is the {\it hyperfine anomaly}. The
ratio $\mu_1/\mu_2$ can be determined from hyperfine experiments on
the two free isotopes/isomers in a known externally applied magnetic
field. Comparison with the ratio as determined from experiments with
the isotopes/isomers incorporated in solids or molecules provides the
value for $\Delta$, which can be as large as 2~\% for heavy elements
like $^{185.187}$Re \cite{BohrWeisskopf_Re}. $\Delta$ is nonzero
because electrons that penetrate the nucleus do not interact with a
point nucleus magnetic moment but with the spatial distribution of the
magnetic moment over the nuclear volume. This slightly affects the
effective hyperfine field.  Therefore, the hyperfine anomaly is
sensitive to the details of nuclear structure, and can be used to test
theoretical nuclear models.

In the same way the $\delta$ from Eq.~(\ref{eq-ratio}) -- which can be called
in analogy the {\it quadrupole anomaly} -- probes details of the nuclear
charge distribution by electrons that penetrate into the nuclear volume. From
Eq.~(\ref{eq-ratio}), it can be seen that $\delta$ is sensitive to the
electronic quantities, $n_{ {zz}}$ and $V_{ {zz}}$, and the {\it difference}
between the squared monopole radii of the two isotopes/isomers that are
involved.

\begin{figure}[t]
\begin{center}
\includegraphics[width=\columnwidth]{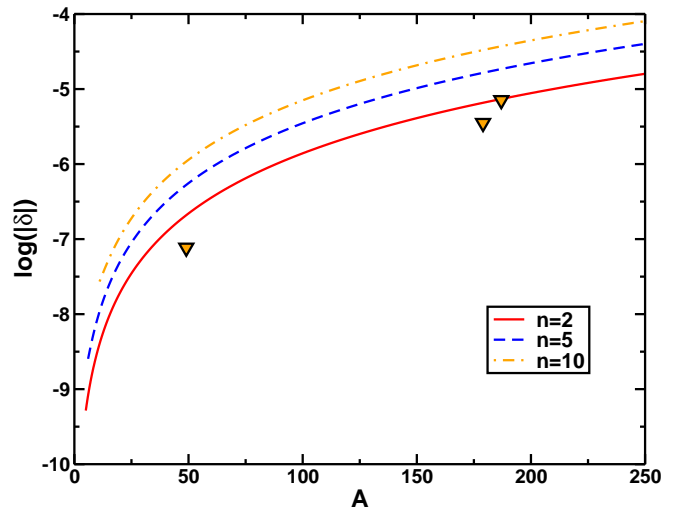}
\end{center}
\caption{The logarithm of the quadrupole anomaly $\delta$ as a
  function of the mass number $A$, as given by
  Eq.~(\ref{eq-trenddelta}). The value of $n$ indicates the mass
  number difference between the heaviest ($A$) and lightest ($A-n$)
  isotope. The curves for $n=2$ (full red line), $n=5$ (blue dashed
  line) and $n=10$ (yellow dot-dashed line) are shown. For $^{47}$Ti,
  $^{179}$Hf and $^{187}$Re (all $n=2$) $\log|\delta|$ can be calculated
  explicitly from Tab.~\ref{QandQStable} (orange triangles).}
\label{fig-trenddelta}
\end{figure}

\begin{table*}[t]
\caption{Ratios of experimental quadrupole coupling constants for two
  different isotopes in two different diatomic molecules, collected
  from the literature. Only cases where the error bar on this ratio
  has been determined directly from the fit to the experimental data
  are reported (this error bar can be slightly different from what one
  would obtain using the error bars on the individual frequencies, see
  the discussion in Ref.~\cite{Cederberg2006a}). The experimental
  value of the EFG (in $10^{21}$V/m$^2$) and the estimated value of
  $\delta$ (Eq.~(\ref{eq-trenddelta})) are given as well.}
\label{ratioTab}
\begin{center}
 \begin{tabular*}{\linewidth}{l@{\extracolsep\fill}lllll}
\hline
\hline
\\[-0.25cm]
Molecules & Isotopes & $\nu_{na}/\nu_{ma}$ & $V_{zz}^{exp}$ & $\left|\delta\right|$ & Ref.
\\[.15cm]
\hline
\\[-0.35cm]
$^6$Li$^{19}$F, $^7$Li$^{19}$F & $^6$Li/$^7$Li & 0.020161
$\pm$ 0.000013 & -0.44 & $5.9\,\cdot\,10^{-10}$ &\cite{Cederberg1998}
\\
$^6$Li$^{127}$I, $^7$Li$^{127}$I & $^6$Li/$^7$Li & 0.02028{\phantom{1}}
$\pm$ 0.00014{\phantom{1}} & -0.18 & &\cite{Cederberg2005}
\\
\\[-0.15cm]
$^{41}$K$^{19}$F, $^{39}$K$^{19}$F & $^{41}$K/$^{39}$K &1.217699{\phantom{5}}
$\pm$ 0.000055 & -5.6 & $1.3\,\cdot\,10^{-7}$ &\cite{Paquette1988}
\\
$^{41}$K$^{127}$I, $^{39}$K$^{127}$I & $^{41}$K/$^{39}$K &1.2174935
$\pm$ 0.0000099 & -3.0 & &\cite{Cederberg2008}
\\
\\[-0.15cm]
$^{87}$Rb$^{19}$F, $^{85}$Rb$^{19}$F& $^{87}$Rb/$^{85}$Rb& 0.4838301
$\pm$ 0.0000018 & -10.7 & $9.6\,\cdot\,10^{-7}$ &\cite{Cederberg2006a}
\\
$^{87}$Rb$^{35}$Cl, $^{85}$Rb$^{35}$Cl& $^{87}$Rb/$^{85}$Rb& 0.483837{\phantom{1}}
$\pm$ 0.000022{\phantom{2}}& -8.2 & &\cite{Cederberg2006b}
\\[.05cm]
\hline
\hline
\end{tabular*}
\end{center}
\end{table*}

In order to find a general trend and order of magnitude estimate for
$\delta$, we combine the analytical function of
Eq.~(\ref{eq-Z4analytic}) with the fitted function of
Eq.~(\ref{eq-A4}) and the square of Eq.~(\ref{nucradius}) to obtain a
numerical approximation for the electronic part ${n_{ {zz}}}/({14
  \epsilon_0 V_{ {zz}}})$ in Eq.~(\ref{eq-ratio}). By inserting this
and the square of Eq.~(\ref{nucradius}) for two different isotopes in
the definition of $\delta$, the following dependence of $\left| \delta
\right|$ on the isotope mass number emerges:
\begin{eqnarray}
\nonumber
|\delta(A)| &=& 5.46 \cdot 10^{-12} \, A^{3.079} \, (A^{0.588}
-(A-n)^{0.588} ).
\\
\label{eq-trenddelta}
\end{eqnarray}
This expression estimates the order of magnitude of $\delta$ for two
isotopes with mass numbers $A$ and $A-n$. Curves for $\log|\delta(A)|$
for $n=2,\,5$ and~10 are shown in Fig.~\ref{fig-trenddelta}. We
observe that the quadrupole anomaly strongly increases with $A$ (or
$Z$), due to the increase of $n_{zz}$. Mass number differences of 10
yield a value for $\delta$ that is an order of magnitude larger than
mass number differences of 2. For the 3 elements in
Tab.~\ref{QandQStable} for which information for 2 isotopes is
provided, Eq.~(\ref{eq-trenddelta}) can be compared by values obtained
by filling out the quantities of Tab.~\ref{QandQStable} directly into
Eq.~(\ref{eq-ratio}). The values are shown by the orange triangles in
Fig.~\ref{fig-trenddelta} and correspond to the red fit ($n=2$).  This
comparison shows that Eq.~(\ref{eq-trenddelta}) is within one order of
magnitude indeed a good estimate for $\delta$. The experimentally
achievable accuracy of quadrupole moment ratios is of the order of
10$^{-6}$ (see Tab.~\ref{ratioTab}). This means that for many isotopes
the presence of $\delta$ affects the experimental values.

Unfortunately, whereas in Bohr-Weisskopf experiments the unperturbed
ratio $\mu_1/\mu_2$ can be determined from experiments on free nuclei
in an externally applied magnetic field, this is not possible for
quadrupole interaction measurements: electric-field gradients that can
be generated by man-made devices are too small to allow meaningful
quadrupole interaction measurements~\cite{HakenWolf}. Therefore, a
slightly different method has to be used. One could perform 4
quadrupole interaction experiments on two isotopes (`$m$' and `$n$')
of the same element, each of them being part of two different
molecules (`$A$' and `$B$'). For instance, $^{m}X$ in $^{m}XA$ and
$^{m}XB$ molecules, and $^{n}X$ in $^{n}XA$ and $^{n}XB$
molecules. This yields four experimental frequencies $\nu_{m}^A$,
$\nu_{n}^A$, $\nu_{m}^B$ and $\nu_{n}^B$. By applying
Eq.~(\ref{eq-ratio}) twice, it can be seen that the NQCC ratios are
not necessarily identical to each other for the two different
molecules, with the difference being determined by $n_{zz}/V_{zz}$:
\begin{eqnarray}
\frac{\nu_{m}^A}{\nu_{n}^A} & = & \frac{Q_m}{Q_n} \left( 1 + \frac{n_{
    {zz}}^A}{14 \epsilon_0 V_{ {zz}}^A} \left( a_n^2 - a_m^2 \right)
\right) \label{eq-ratio1} \\
\frac{\nu_{m}^B}{\nu_{n}^B} & = & \frac{Q_m}{Q_n} \left( 1 + \frac{n_{
    {zz}}^B}{14 \epsilon_0 V_{ {zz}}^B} \left( a_n^2 - a_m^2 \right)
\right).
\label{eq-ratio2}
\end{eqnarray}
As long as the quadrupole shift ($\propto n_{zz}$) does not play a
significant role, the two experimental frequency ratios at the
left-hand side are within their error bars identical to each
other. If, however, the quadrupole shift would be large enough, these
two experimental frequency ratios would differ from each other. {\it
  This is a completely experimental procedure to detect the presence
  of the quadrupole shift effect.} Tab.~\ref{ratioTab} lists a
collection of experimental NQCC-ratios in diatomic molecules
determined for three such sets of 4 experiments, which gives an
impression of the experimental accuracy that can be achieved. The
estimated order of magnitude for $\left|\delta\right|$
(Eq.~(\ref{eq-trenddelta})) is given too. For none of these cases,
$\delta$ is expected to be large enough to affect the experimental
ratios. Tab.~\ref{ratioTab} combined with Fig.~\ref{fig-trenddelta}
suggests that if the best experimental accuracies of 10$^{-6}$ can be
achieved for isotopes with A$\ge$150, then the influence of $\delta$
could be observed. The heavier the element and the larger the
size-differences between the two isotopes, the more likely large
$\delta$-values are. Interestingly enough, the quadrupole coupling
constant ratios for the two K isotopes in the KF and KI molecules
differ from each other in the 4$^{th}$ digit, and this difference is
an order of magnitude larger than the experimental error bars. Given
the estimate for $\delta$, the quadrupole shift is expected to give an
effect in the 7$^{th}$ digit at best. It is therefore unlikely that
this set of K-experiments represents an experimental observation of
the quadrupole shift (it could be due to one of the other effects
discussed in App.~\ref{appendix}, or due to an experimental
problem). Nevertheless, it would be interesting to perform similar
experiments with the same accuracy for heavier elements, where
$\delta$ is expected to be larger.

One step further is to solve the system of the two
equations~(\ref{eq-ratio1}) and~(\ref{eq-ratio2}) for the unknown
quantities $Q_m/Q_n$ and $\left(a^2_n-a^2_m\right)$:
\begin{eqnarray}
\frac{Q_m}{Q_n} & = & \frac{\frac{\nu_{mb}}{\nu_{nb}} \frac{n_{ {zz}}^a}{14
    \epsilon_0 V_{ {zz}}^a} - \frac{\nu_{ma}}{\nu_{na}} \frac{n_{ {zz}}^b}{14
    \epsilon_0 V_{ {zz}}^b}}{\frac{n_{ {zz}}^a}{14 \epsilon_0 V_{ {zz}}^a} -
  \frac{n_{ {zz}}^b}{14 \epsilon_0 V_{ {zz}}^b}} \label{eq-Qratio-solution}
\\ a_n^2 - a_m^2 & = & \frac{\frac{\nu_{ma}}{\nu_{na}} -
  \frac{\nu_{mb}}{\nu_{nb}}}{\frac{\nu_{mb}}{\nu_{nb}} \frac{n_{ {zz}}^a}{14
    \epsilon_0 V_{ {zz}}^a} - \frac{\nu_{ma}}{\nu_{na}}\frac{n_{ {zz}}^b}{14
    \epsilon_0 V_{ {zz}}^b}}. \label{eq-adiff-solution}
\end{eqnarray}
All quantities at the right-hand side of these equations can either be
measured or calculated, such that the quantities at the left-hand side
are effectively determined by a combination of experiment and
theory. Clearly, this is a game with very small numbers. The
difference between the two frequency ratios in the numerator of
Eq.~(\ref{eq-adiff-solution}) is of the same order of magnitude as the
$\delta$ in Eq.~(\ref{eq-ratio}): 10$^{-5}$ for heavy elements. The
same considerations as in Sec.~\ref{subsec-QtQ-method} apply here: an
extreme accuracy in experiments as well as in calculations is needed
in order to get to a reliable conclusion. Furthermore, the procedure
as described here can be disturbed by the presence of a few other
small quadrupole-like effects that are discussed in
Sec.~\ref{sec-other} and App.~\ref{appendix}.

%
%
%
%
%
%
%
%

\section{Conclusions and outlook}
In this work, we described how electron penetration in the nuclear
volume leads to the {\it quadrupole shift}: a small perturbation of
the regular quadrupole interaction, which depends on the second
derivative of the electron charge density at the nucleus ($n_{zz}$),
as well as on the size and shape of the nucleus ($\tilde{Q}$). An
explicit expression for the quadrupole shift that can be implemented
in a band structure code was derived, and DFT calculations were
performed for a set of crystalline materials. It was shown that
meaningful numerical values for the quadrupole shift can be obtained
only for fully relativistic calculations that take a finite nucleus
into account. Therefore, the quadrupole shift is one of the few cases
where the commonly used scalar-relativistic approximation is
definitely insufficient.

The quadrupole shift is a small effect. Its order of magnitude appears
to be related in the first place to the atomic number $A$ of the
element under consideration, and to a lesser extent to the crystal
structure (Fig.~\ref{fig-logtrend}). This is predominantly due to the
way how $n_{zz}$ depends on $Z$. The quadrupole shift is orders of
magnitude smaller than the regular quadrupole interaction for most
elements, but can reach 1\% to perhaps 10\% near the actinide region.

We have pointed out how the quadrupole shift can play a role in a more
accurate determination of quadrupole moments and quadrupole moment
ratios. The comparison of two accurately measured quadrupole coupling
constant ratios provides a purely experimental way to observe the
presence of the quadrupole shift. For suitable cases, the required
experimental accuracy can be reached by e.g.\ molecular beam
spectroscopy. With further advances in the absolute accuracy of {\it
  ab~initio} calculations for $n_{zz}$ and $V_{zz}$, awareness of the
existence of the quadrupole shift will help to extract more precise
nuclear information from quadrupole coupling experiments.

Suggestions for further work are at the conceptual, computational as
well as on the experimental level. Conceptual: it remains to be
understood which features of the electron density are responsible for
the observed $Z$-dependence of $n_{zz}$ and for the dependence of
$n_{zz}$ for a given element on the crystal structure. Understanding
those mechanisms would help to single out situations where the
quadrupole shift is maximized. Computational: in the present work,
only DFT calculations for solids were performed, whereas the most
accurate experiments are available for molecules. DFT for molecules is
not likely to provide very accurate results, but quantum chemical
calculations can do much better in this respect. It would be
interesting to examine for instance the value of the quadrupole shift
for heavy elements in a set of molecules. Experimental: sets of 4
quadrupole coupling experiments as in Tab.~\ref{ratioTab}, done for
heavy elements and with high accuracy, provide a way to observe the
presence of the quadrupole shift experimentally. It would be most
efficient to make first a computational study, to identify among those molecules that
are experimentally most easily accessible the ones where a large quadrupole shift is most likely.

\section*{Acknowledgments}

This work has grown through numerous discussion with many people, and
we warmly acknowledge the colleagues who shared their knowledge on
experiment, theory, physics, chemistry, mathematics, literature and
history (in alphabetical order): Tim Bastow (CSIRO, Australia), Peter
Blaha (TU Wien), Jim Cederberg (St. Olaf College, Northfield),
%
%
Frank Haarmann (MPI CPfS, Dresden), Heinz Haas (HMI, CERN), Ralph
Haberkern (RWTH, Aachen), Gerda Neyens (K.U.Leuven), Ingo Opahle
(CMT, Frankfurt a. M.), Pekka Pyykk\"o (University of Helsinki),
Karlheinz Schwarz (TU Wien), Peter Schwerdtfeger (Massey University,
Auckland), Nathal Severijns (K.U.Leuven), Christian Thierfelder
(Massey University, Auckland),
%
%
and Reiner Vianden (Bonn).

Furthermore, the International Max Planck Research School for
``Dynamical Processes in Atoms, Molecules and Solids" (Dresden) and
and the {\it Fonds voor Wetenschappelijk Onderzoek - Vlaanderen} (FWO)
of Flanders (Belgium) (FWO-project G.0501.07) are acknowledged for
financial support.

\appendix

%
%
%
%
%
%
%
%
%
%
%

\section{Other small perturbations to the quadrupole interaction}
\label{appendix}
As announced in Sec.~\ref{sec-other}, this appendix discusses several
other small quadrupole-like effects that might be of similar magnitude
of the quadrupole shift.

\subsection{Second order effects of magnetic origin}
\label{subsec-pseudoQI}
Van~Vleck, Rabi, Foley and
Ramsey~\cite{Kellog1940,Foley1947,Ramsey1953} discussed half a century
ago a {\it pseudo quadrupole interaction} in molecules that has a
magnetic origin. This has been later elaborated upon by P.~Pyykk\"o,
especially for the case of
metals~\cite{Pyykko1970,Pyykko1971}. Mathematically, this pseudo
quadrupole interaction arises in the same way as the quadrupole shift
when the latter is derived by second order perturbation theory (see
the end of Sec.~\ref{Hamiltonians}). The small perturbing Hamiltonians
are now the ones that give rise to the magnetic hyperfine field: the
nuclear spin/electron orbit Hamiltonian ($\hat{\mathcal{H}}_1$), the
nuclear spin/electron spin dipole-dipole Hamiltonian
($\hat{\mathcal{H}}_2$) and the Fermi contact Hamiltonian
($\hat{\mathcal{H}}_3$). In second order perturbation, the square of
$\hat{\mathcal{H}}_1$, the square of $\hat{\mathcal{H}}_2$ and the
cross-term of $\hat{\mathcal{H}}_2$ and $\hat{\mathcal{H}}_3$ contain
an $\hat{I}_z^2$ operator that gives rise to a quadrupole-like
interaction (compare to Eq.~(\ref{eq-regular-QI-hamiltonian}),
provided axial symmetry ($\eta=0$) is present). Such a term is
included even in first order in $\hat{\mathcal{H}}_1$. These pseudo
quadrupole interactions were shown to be at the level of
10$^{-4}$-10$^{-6}$ of the regular quadrupole interaction in
molecules~\cite{Foley1947,Ramsey1953,Pyykko1970}, and reach in
favorable cases up to 1\% in metals~\cite{Pyykko1970}. The values in
molecules are therefore of the same order of magnitude as the
quadrupole shift (Fig.~\ref{fig-logtrend} and Tab.~\ref{QandQStable}).

Strictly spoken, these quadrupole-like contributions have a different
status than the quadrupole shift in Sec.~\ref{sec-formalism}. The
quadrupole shift Hamiltonian (Eq.~(\ref{quadruSH})) has exactly the
same structure as the quadrupole interaction Hamiltonian
(Eq.~(\ref{eq-regular-QI-hamiltonian})), and they are therefore
completely indistinguishable from each other. The quadrupole-like
interactions discussed in the present section have in the first place
a $\hat{I}_z^2$ dependence which splits the nuclear levels in a
quadrupole-like manner as long as the environment has axial symmetry
($\eta=0$). The important case of linear molecules has this
symmetry. In less symmetrical environments, one could in principle
distinguish between these quadrupole-like interactions and the
quadrupole shift. As this symmetry breaking is itself a small effect,
however, such considerations are not expected to be of much practical
value.

For completeness, we mention two other sources of quadrupole-like
interactions, which are believed~\cite{Pyykko1971} to be even smaller
than the previously described ones: the influence of an external
magnetic field~\cite{Ganiel1968,Kamal1970} and nuclear
polarization~\cite{Bonczyk1970}.

%
%
%
%
%
%
%
%
%
%
%

\subsection{The isotopologue anomaly}
\label{TheIA}

\begin{table*}[t]
\begin{center}
\begin{tabular*}{\linewidth}{l@{\extracolsep\fill}lrrrrrr}
\hline
\hline
molecule A & molecule B & $\nu_{Q,A}$ & $\nu_{Q,B}$ & $\Delta \nu$ & $\Delta \nu_{rel}$ & $\Delta m_{rel}$ & Ref.
\\
           &            &   (MHz)          &   (MHz)    & (kHz) & (\%) & (\%) & \\
\hline
$^1$H{\bf $^{81}$Br} & $^2$H{\bf $^{81}$Br} & 447.9(14) & 443.363(105) & 4537(1500)* & 1.023\% & 50.0\% & \cite{DeLucia1971} \\
$^1$H{\bf $^{79}$Br} & $^2$H{\bf $^{79}$Br} & 535.4(14) & 530.648(74) & 4752(1500)* & 0.896\% & 50.0\% & \cite{DeLucia1971} \\
$^1$H{\bf $^{37}$Cl} & $^2$H{\bf $^{37}$Cl} & -53.436(95) & -53.037(113) & -399(200)* & 0.752\% & 50.0\% & \cite{DeLucia1971} \\
$^1$H{\bf $^{35}$Cl} & $^2$H{\bf $^{35}$Cl} & -67.800(95) & -67.417(98) & -383(200)* & 0.568\% & 50.0\% & \cite{DeLucia1971} \\
$^1$H{\bf $^{127}$I} & $^2$H{\bf $^{127}$I} & -1828.059(51) & -1823.226(54) & -4833(100)* & 0.265\% & 50.0\% & \cite{Chance1993,Varberg1998} \\
$^2$H{\bf $^{35}$Cl} & $^3$H{\bf $^{35}$Cl} & -67.417(98) & -67.0(6) & -417(700)\phantom{*} & 0.622\% & 33.3\% & \cite{DeLucia1971} \\
$^2$H{\bf $^{79}$Br} & $^3$H{\bf $^{79}$Br} & 530.648(74) & 530(2) & 648(2100)\phantom{*} & 0.122\% & 33.3\% & \cite{DeLucia1971} \\
$^2$H{\bf $^{81}$Br} & $^3$H{\bf $^{81}$Br} & 443.363(105) & 443(2) & 363(2100)\phantom{*} & 0.082\% & 33.3\% & \cite{DeLucia1971} \\
$^2$H{\bf $^{37}$Cl} & $^3$H{\bf $^{37}$Cl} & -53.037(113) & -53.0(6) & -37(700)\phantom{*} & 0.070\% & 33.3\% & \cite{DeLucia1971} \\
$^6$Li{\bf $^{127}$I} & $^7$Li{\bf $^{127}$I} & -194.33834(20) & 194.35241(20) & 14.07(40)* & 0.007\% & 14.3\% & \cite{Cederberg2005} \\
$^{35}$Cl{\bf $^{45}$Sc} & $^{37}$Cl{\bf $^{45}$Sc} & 68.2067(29) & 68.2062(29) & 0.5(6.0)\phantom{*} & 0.000\% & 5.4\% & \cite{Lin2000} \\
$^{39}$K{\bf $^{127}$I} & $^{41}$K{\bf $^{127}$I} & -85.471138(7) & -85.471721(12) & 0.583(20)* & 0.001\% & 4.8\% & \cite{Cederberg2008} \\
$^{63}$CuOC-{\bf $^{127}$I} & $^{65}$KOC-{\bf $^{127}$I} & -593.465(9) & -593.485(10) & 20(20)\phantom{*} & 0.003\% & 3.1\% & \cite{Batten2006} \\
$^{79}$Br{\bf $^{39}$K} & $^{81}$Br{\bf $^{39}$K} & -5.032957(9) & -5.032957(9) & 0.000(20)\phantom{*} & 0.000\% & 2.5\% & \cite{Cederberg2008} \\
$^{79}$Br{\bf $^{45}$Sc} & $^{81}$Br{\bf $^{45}$Sc} & 65.2558(32) & 65.2597(38) & -3.9(7.0)\phantom{*} & 0.000\% & 2.5\% & \cite{Lin2000b}
\\
\hline
\hline
\end{tabular*}
\caption{\label{tab-IA} The isotopologue anomaly for a set of diatomic
  molecules. First two columns: the molecules with their isotopes --
  the isotope for which the NQCC is measured is put in bold. Second
  two columns: the NQCC in MHz. $\Delta \nu$: the difference between the preceding two columns (kHz) -- cases where the error bar allows to conclude the
  difference is not zero, are labeled by a ``*'". $\Delta \nu_{rel}$:
  relative frequency difference (\%). $\Delta m_{rel}$: relative change
  in atomic mass number for the neighboring isotope (\%).}
\end{center}
\end{table*}

Through high-precision molecular beam experiments, Cederberg~{\it et
  al.} have drawn attention to the fact that the quadrupole
interaction at nucleus B in an AB diatomic molecule slightly depends
on which isotope is taken for element A. For the
isotopologues\footnote{Most of the literature in this context uses the
  word `isotopomers' to refer to (e.g.) the pair $^7$Li$^{127}$I and
  $^6$Li$^{127}$I. According to the 1994 IUPAC
  recommendations~\cite{IUPAC1994}, `isotopologues' is the better term
  for this.} $^7$Li$^{127}$I and $^6$Li$^{127}$I, this {\it
  isotopologue anomaly} is 0.007\% of the regular quadrupole
interaction at $^{127}$I (an absolute shift of
14~kHz)~\cite{Cederberg2005}. For $^{41}$K$^{127}$I and
$^{39}$K$^{127}$I, the relative effect at $^{127}$I was 10 times
smaller (0.0007\%, absolute shift of 0.6~kHz)~\cite{Cederberg2008},
while for $^{39}$K$^{81}$Br and $^{39}$K$^{79}$Br there was no effect
found at all on $^{39}$K~\cite{Cederberg2008}. The origin of the
isotopologue anomaly is not
understood~\cite{Cederberg2005,Cederberg2008}, but from the literature
review we present in Tab.~\ref{tab-IA}, a correlation between the
relative value of the isotopologue anomaly and the relative mass
number change for the A-isotope is clearly present. The 50\% relative
mass change between hydrogen and deuterium results in an isotopologue
anomaly of 0.2-1.0\%. It tends to be lower for the 33\% mass change
between deuterium and tritium (0.1-0.6\%), although the error bars
prevent unambiguous conclusions. Much smaller but definitely non-zero
frequency differences are observed for LiI and KI as well (0.007\% and
0.001\%). Tab.~\ref{tab-IA} suggests that these isotopologue anomalies
tend to become undetectably small for relative mass number changes
below 5\%.

The isotopologue anomalies as presented in Tab.~\ref{tab-IA} were
obtained as the difference between ($\nu$=0,~$J$=0)-terms in the
vibrational/rotational expansion of the quadrupole coupling (see
e.g.~Eq.~(19) in Ref.~\cite{Cederberg1999}). This lowest order term is
not exactly equal to the static quadrupole coupling constant at the
equilibrium internuclear separation, due to the presence of an
additional constant (the $\alpha B^2$ term in Eq.~(18) of
Ref.~\cite{Cederberg1999}) which is mass-dependent and therefore
isotope-dependent. This $\alpha B^2$ term could therefore be an
obvious candidate to explain the observed frequency
difference. However, Cederberg~{\it et~al.} have shown for CsF that
$\alpha B^2$ is negligible~\cite{Cederberg-CsF}, because it is an
order of magnitude smaller than the ($\nu$=2, $J$=0)-term, which
itself is experimentally known to be small. We verified that the same
argument holds true for LiI ($\Delta
m_{rel}=14\%$)~\cite{Cederberg1999} and HI~($\Delta
m_{rel}=50\%$)~\cite{Matsushima1991,Chance1993}. Therefore, it is safe
to conclude that the larger as well as the smaller frequency
differences in Tab.~\ref{tab-IA} are not significantly influenced by
the $\alpha$B$^2$ term and represent a real difference between two
static quadrupole coupling constants, a difference of which the origin
remains to be understood.

Isotopologue anomalies in the kHz region can be of the same order of
magnitude as the quadrupole shift $\nu_{QS}$. Their existence puts
further limitations on the numerical information that can be extracted
from a comparison of experimental quadrupole coupling constants and
{first-principles} calculations. Indeed, the {\it
  first-principles} values for $V_{zz}$ and $n_{zz}$ that appear for
instance in Eqs.~(\ref{eq-Qratio-solution})
and~(\ref{eq-adiff-solution}) can only be calculated for specified
{\it elements} in the molecule, not for the {\it isotopes}. On the
other hand, the purely experimental determination of the presence of a
quadrupole shift by 4 NQCC measurements as in Eqs.~(\ref{eq-ratio1})
and~(\ref{eq-ratio2}) is not disturbed by the isotopologue anomaly, as
long as one makes sure that the isotopes for A and B in the
$^{(m,n)}$XA and $^{(m,n)}$XB molecules remain identical in all 4
cases.

%
%
%
%
%
%
%
%
%
%
%

\subsection{Temperature and vibrations}
The entire discussion so far implicitly assumed static molecules or
crystals (0~K and no zero point vibrations). At non-zero temperatures,
vibrational states will be populated, and in molecules rotational
states as well. These will influence the electric-field gradient and
therefore the quadrupole interaction. The effect is in the range of
1-10\%, and should therefore certainly be taken into account in
high-precision studies. In molecules, this effect can be described
with high accuracy using a Schlier-Dunham
treatment~\cite{Dunham1932,Schlier1961,Cederberg1999,Cederberg2005},
and quadrupole coupling experiments are routinely analyzed according
to this formalism~\cite{Matsushima1991, Chance1993, Cederberg1999,
  Lin2000, Lin2000b, Cederberg2005}. Similar studies in solids are
rare -- see e.g.~Ref.~\cite{Torumba2006} for hcp-Cd, where a
contribution of 1.6\% due to zero-point vibrations was found. \\


\end{document}